\documentclass[a4paper,11pt,reqno]{article}
\usepackage{a4wide}
\usepackage{amsmath,amsfonts,amssymb}
\usepackage[english]{babel}
\usepackage{soul}
\usepackage{nicefrac}
\usepackage[mathscr]{euscript}
\usepackage{setspace}
\usepackage{datetime}
\usepackage[nosort]{cite}
\usepackage{mathrsfs}
\usepackage[T1]{fontenc}

\usepackage{txfonts}
\usepackage{kpfonts}
\usepackage{upgreek}

\usepackage{comment}

\usepackage{mathpazo}
\usepackage{graphicx}
\usepackage[breaklinks=true]{hyperref}

\numberwithin{equation}{section}

\hypersetup{
			colorlinks=true,
			urlcolor=blue,
			citecolor=magenta,
			linkcolor=blue,
     }

\let\oldsqrt\sqrt
\def\sqrt{\mathpalette\DHLhksqrt}
\def\DHLhksqrt#1#2{%
\setbox0=\hbox{$#1\oldsqrt{#2\,}$}\dimen0=\ht0
\advance\dimen0-0.2\ht0
\setbox2=\hbox{\vrule height\ht0 depth -\dimen0}%
{\box0\lower0.4pt\box2}}

\newcommand{\RNum}[1]{\uppercase\expandafter{\romannumeral #1\relax}}

\author{
  \begin{minipage}{.97\linewidth}
    \vspace{1cm}
       \begin{center}
      \begin{small}
               \textbf{Luca Ciambelli},$^1$  
               \textbf{Charles Marteau},$^1$ 
             \textbf{Anastasios C. Petkou},$^{2,3}$ 
                     \\
     \textbf{P. Marios Petropoulos}$^1$ and 
      \textbf{Konstantinos Siampos}$^{3,4}$
              \end{small}
    \end{center}
    \vspace{0.5cm}
      \hspace{2.4cm}\begin{minipage}{.7\linewidth}
\begin{center}     {\it \begin{footnotesize}
\hbox{\kern-1.8cm\vbox{\vskip0cm
 \begin{itemize}
               \item[$^1$]CPHT -- Centre de Physique Th\'eorique\\ 
        Ecole Polytechnique, CNRS UMR 7644\\
        Universit\'e Paris--Saclay\\
        91128 Palaiseau Cedex, France
\vskip0.3cm
      \end{itemize}}
\kern-3cm\vbox{
\begin{itemize}
 \item[$^2$]Department of Physics\\ 
  Institute of Theoretical Physics\\
  Aristotle University of Thessaloniki\\ 
  54124 Thessaloniki, Greece
      \end{itemize}
      \vskip0.05cm
}}
     \end{footnotesize}}
\end{center}
    \end{minipage}
    \vspace{0.5cm}\begin{minipage}{.7\linewidth}
\begin{center}     
{\it \begin{footnotesize}
\hbox{\kern0.6cm\vbox{\vskip0cm
 \begin{itemize}
              \item[$^3$] Theoretical Physics Department\\
CERN\\ 
1211 Geneva 23, Switzerland
\vskip0.45cm
      \end{itemize}}
\kern-3cm\vbox{
\begin{itemize}
 \item[$^4$] Albert Einstein Center for Fundamental Physics\\
Institute for Theoretical Physics\\ 
University of Bern\\
Sidlerstrasse 5, 3012 Bern, Switzerland
      \end{itemize}\vskip0.05cm
}
}
     \end{footnotesize}}
\end{center}
     \end{minipage}
  \end{minipage}
}

\title{\vspace{1.5cm}
 \boldmath 
    \textbf{Flat holography and Carrollian fluids}
  \unboldmath
}

\date{}

\begin{document}

\begin{titlepage}
\maketitle
\thispagestyle{empty}

 \vspace{-14.cm}
  \begin{flushright}
  CPHT-RR049.082017\\
CERN-TH-2017-229
  \end{flushright}
 \vspace{11.7cm}

\begin{center}
\textsc{Abstract}\\  
\vspace{0.5cm}	
\begin{minipage}{1.0\linewidth}
We show that a holographic description of four-dimensional asymptotically locally flat spacetimes is reached { smoothly} from the zero-cosmological-constant limit of anti-de Sitter holography. { To this end, we use the derivative expansion of fluid/gravity correspondence.}  From the boundary perspective, the vanishing of the bulk cosmological constant appears as the zero velocity of light limit. This sets how Carrollian geometry emerges in flat holography. The new boundary data are a two-dimensional spatial surface, identified with
the null infinity of the bulk Ricci-flat spacetime, accompanied with a Carrollian time and equipped with a Carrollian structure, plus the dynamical observables of a conformal Carrollian fluid. These are the energy, the viscous stress tensors and the heat currents, whereas the Carrollian geometry is gathered by a two-dimensional spatial metric, a frame connection and a scale factor. The reconstruction of Ricci-flat spacetimes from Carrollian boundary data is conducted with a flat derivative expansion, resummed in a closed form in Eddington--Finkelstein gauge
under further integrability conditions inherited from the ancestor anti-de Sitter set-up. These conditions are hinged on a duality relationship among fluid friction tensors and Cotton-like geometric data. 
We illustrate these results in the case of conformal Carrollian perfect fluids and Robinson--Trautman viscous hydrodynamics. The former are dual to the asymptotically flat Kerr--Taub--NUT 
family, while the latter leads to the homonymous class of algebraically special Ricci-flat spacetimes. 

\end{minipage}
\end{center}


\end{titlepage}

\onehalfspace

\begingroup
\hypersetup{linkcolor=black}
\tableofcontents
\endgroup
\noindent\rule{\textwidth}{0.6pt}

\section{Introduction}

Ever since its conception, there have been many attempts to extend the original holographic anti-de Sitter correspondence along various directions, including asymptotically flat or de Sitter bulk spacetimes. Since the genuine microscopic correspondence based on type IIB string and maximally supersymmetric Yang--Mills theory is deeply rooted in the anti-de Sitter background, phenomenological extensions such as fluid/gravity correspondence have been considered as more promising for reaching a flat spacetime generalization.

The mathematical foundations of holography are based on the existence of the Fefferman--Graham expansion for asymptotically anti-de Sitter Einstein spaces \cite{PMP-FG1, PMP-FG2}.
Indeed, on the one hand, putting an asymptotically anti-de Sitter Einstein metric in the Fefferman--Graham gauge allows to extract the two independent boundary data \emph{i.e.} the boundary metric and the conserved boundary conformal energy--momentum tensor. On the other hand, given a pair of suitable boundary data the Fefferman--Graham expansion makes it possible to reconstruct, order by order, an Einstein space.

More recently, fluid/gravity correspondence has provided an alternative to Fefferman--Graham, known as derivative expansion \cite{Bhattacharyya:2007, Hubeny:2011hd,Haack:2008cp, Bhattacharyya:2008jc}.  It is inspired from the fluid derivative expansion (see \emph{e.g.} \cite{Kovtun:2012rj, Romatschke:2009im}), and is implemented in Eddington--Finkelstein coordinates. The metric of an Einstein spacetime is  expanded in a light-like direction and the information on the boundary fluid is made available in a slightly different manner, involving explicitly a velocity field whose derivatives set the order of the expansion. Conversely, the  boundary fluid data, including the fluid's congruence, allow to reconstruct an exact bulk Einstein spacetime.

Although less robust mathematically, the derivative expansion has several advantages over Fefferman--Graham. Firstly,  under some particular conditions it can be resummed leading to algebraically special Einstein spacetimes in a closed form \cite{Caldarelli:2012cm, Mukhopadhyay:2013gja, Petropoulos:2014yaa, Gath:2015nxa, Petropoulos:2015fba, Petkou:2015fvh}. Such a resummation is very unlikely, if at all possible, in the context of Fefferman--Graham. Secondly, boundary geometrical terms appear packaged at specific orders in the derivative expansion,
which is performed in Eddington--Finkelstein gauge. These terms feature precisely whether the bulk is asymptotically globally or locally anti-de Sitter. 
Thirdly, and contrary to Fefferman--Graham again, the derivative expansion admits a consistent limit of vanishing scalar curvature. Hence it appears to be applicable to Ricci-flat spacetimes and emerges as a valuable tool for setting up flat holography. Such a smooth behaviour is not generic, as in most coordinate systems switching off the scalar curvature for an Einstein space leads to plain Minkowski spacetime.\footnote{This phenomenon is well known in supergravity, when studying the gravity decoupling limit of scalar manifolds. For this limit to be non-trivial, one has to chose an appropriate gauge (see \cite{Antoniadis:2016kde,Alexandrov:2017mgi} for a recent discussion and references).}

The observations above suggest that 
it is relevant to wonder whether a Ricci-flat spacetime admits a dual fluid description.
This can be recast into two sharp questions:
\begin{enumerate}
\item Which surface $\mathscr{S}$ would replace the AdS conformal boundary $\mathscr{I}$, and what is the geometry that this new boundary should be equipped with?
\item Which are the  degrees of freedom hosted by $\mathscr{S}$ and succeeding the relativistic-fluid energy--momentum tensor, and what is the dynamics these degrees of freedom obey?
\end{enumerate}

Many proposals have been made for answering these questions.  Most of them were inspired by the seminal work \cite{Damourpaper, damour1979quelques}, where Navier--Stokes equations were shown to capture the dynamics of black-hole horizon perturbations. This result is taken as the crucial evidence regarding the deep relation between gravity, without cosmological constant, and fluid dynamics. 

A more recent approach has associated Ricci-flat spacetimes in $d+1$ dimensions with $d$-dimensional fluids \cite{Ske0, Stro1, Ske1, Ske2, Ske3, Tay}.  This is based on the observation that the Brown--York energy--momentum tensor on a Rindler hypersurface of a flat metric has the form of a perfect fluid \cite{oz}. 
In this particular framework, one can consider a non-relativistic limit, thus showing that the Navier--Stokes equations coincide with Einstein's equations on the Rindler hypersurface. Paradoxically, 
 it has simultaneously been argued that all information can be stored in a relativistic $d$-dimensional fluid. 
 
 { Outside the realm of fluid interpretation, and on the more mathematical side of the problem, some solid works regarding flat holography are
\cite{Dappiaggi1, Dappiaggi2, Moretti} (see also \cite{Solodukhin}). The dual theories reside at null infinity emphasizing the importance of the null-like formalisms of \cite{Penrose, Newman, wini}. 
In this line of thought, results where also reached focusing on the expected symmetries, in particular for the specific case of three-dimensional bulk versus two-dimensional boundary \cite{Bagchi:2009my, Bagchi, Bagchi2, Glenn2, Barnich:2012aw,Bagchi:2014iea, Hartong:2015usd}.\footnote{ Reference \cite{Barnich:2012aw} is the first where a consistent and non-trivial $k\to 0$ limit was taken, mapping the entire family of three-dimensional Einstein spacetimes (locally AdS) to the family of Ricci--flat solutions (locally flat).\label{glenn}}
These achievements \emph{are not} unconditionally transferable to four or higher dimensions, and can  
possibly infer inaccurate expectations due to features holding exclusively in three dimensions. } 

The above wanderings between relativistic and non-relativistic fluid dynamics in relation with Ricci-flat spacetimes are partly due to the { incomplete} understanding on the  r\^ole played by the null infinity. On the one hand, it has been recognized that the Ricci-flat limit is related to some 
 contraction of the Poincar\'e algebra \cite{Bagchi:2009my, Bagchi, Bagchi2, Glenn2, Barnich:2012aw,Karch,Izadi}. 
On the other hand, this observation was tempered by a potential confusion among
the Carrollian algebra and its dual contraction, the conformal Galilean algebra, as they both lead to the decoupling of time. { This phenomenon was exacerbated by the equivalence of these two algebras in two dimensions, and has somehow}
obscured the expectations on the nature and the dynamics of the relevant boundary degrees of freedom. { Hence, although the idea of localizing the latter on the spatial surface at null infinity was suggested (as \emph{e.g.} in \cite{Str0,Str1,Str2,Praha}), their description { has often been} accustomed to the relativistic-fluid { or the conformal-field-theory} approaches, based on the revered  energy--momentum tensor and its conservation law.\footnote{This is manifest in the very recent work of Ref. \cite{fareghbal}. }}

From this short discussion, it is clear that the attempts implemented so far follow different directions without clear overlap and common views. Although implicitly addressed in the literature, the above two questions have not been convincingly answered, and the treatment of boundary theories in the zero cosmological constant limit remains nowadays tangled.

In this work we make a precise statement, which clarifies unquestionably the situation.
{ Our starting point is a four-dimensional bulk Einstein spacetime with $\Lambda=-3k^2$, dual to a boundary relativistic fluid. In this set-up, we consider the $k\to 0$ limit, 
which has the following features:  }
\begin{itemize}
\item The derivative expansion is generically well behaved. We will call its limit the \emph{flat derivative expansion}. Under specified conditions it can be resummed in a closed form.
\item Inside the boundary metric, and in the complete boundary fluid dynamics, $k$ plays the r\^ole of \emph{velocity of light}. Its vanishing is thus a \emph{Carrollian limit}.
\item The boundary is the two-dimensional \emph{spatial} surface 
$\mathscr{S}$
emerging as the future null infinity of the limiting Ricci-flat bulk spacetime. It replaces the AdS conformal boundary and is endowed with a \emph{Carrollian geometry i.e.} is covariant under \emph{Carrollian diffeomorphisms}.

\item The degrees of freedom hosted by this surface are captured by a \emph{conformal Carrollian fluid} : energy density and pressure related by a conformal equation of state, heat currents and traceless viscous stress tensors. 
These macroscopic degrees of freedom obey \emph{conformal Carrollian fluid dynamics}. 
\end{itemize}
Any two-dimensional conformal Carrollian fluid hosted by an arbitrary spatial surface $\mathscr{S}$, and obeying conformal Carrollian fluid dynamics on this surface, is therefore mapped onto a Ricci-flat four-dimensional spacetime using the flat derivative expansion. The latter is invariant under boundary Weyl transformations. Under a set of resummability conditions involving the Carrollian fluid and its host $\mathscr{S}$, this derivative expansion allows to reconstruct exactly algebraically special Ricci-flat spacetimes. The results summarized above answer in the most accurate manner the two questions listed earlier.
 
Carrollian symmetry has sporadically attracted attention following the pioneering work or Ref. \cite{Levy}, where the Carroll group emerged as a new contraction of the Poincar\'e group: the ultra-relativistic contraction, dual to the usual non-relativistic one leading to the Galilean group. Its conformal extensions were explored latterly \cite{Duval:2014uoa,Duval:2014uva,Duv,DuvG}, 
showing in particular its relationship to the BMS group, which encodes the asymptotic symmetries of asymptotically flat spacetimes along a null direction \cite{BMS1,BMS2,Glenn,Ash}.\footnote{Carroll symmetry has also been explored in connection to the tensionless-string limit, see \emph{e.g.} \cite{ba,card}.}

It is therefore quite natural to investigate on possible relationships between Carrollian asymptotic structure and flat holography and, by the logic of fluid/gravity correspondence, to foresee the emergence of Carrollian hydrodynamics rather than any other, relativistic or Galilean fluid. Nonetheless searches so far have been oriented towards the near-horizon membrane paradigm, trying to comply with the inevitable BMS symmetries as in \cite{Penna1,Penna2}. The power of the derivative expansion and its flexibility to handle the zero-$k$ limit has been somehow dismissed. This expansion stands precisely at the heart of our method. Its actual implementation requires a comprehensive approach to Carrollian hydrodynamics, as it emanates from the ultra-relativistic limit of relativistic fluid dynamics, made recently available in \cite{CMPPS1}.

The aim of the present work is to provide a detailed analysis of the various statements presented above, and exhibit a precise expression for the Ricci-flat line element as reconstructed from the boundary Carrollian geometry and Carrollian fluid dynamics. As already stated, the tool for understanding and implementing operationally these ideas is the derivative expansion and, under conditions, its resummed version. For this reason, Sec. \ref{sec:resform} is devoted to its thorough description in the framework of ordinary anti-de Sitter fluid/gravity holography. This chapter includes the conditions, stated in a novel fashion with respect to \cite{Gath:2015nxa, Petropoulos:2015fba}, for the expansion to be resummed in a closed form, representing generally an Einstein spacetime of algebraically special Petrov type. 

In  Sec. \ref{sec:Ricci-flat} we discuss how the Carrollian geometry emerges at null infinity and describe in detail conformal Carrollian hydrodynamics following \cite{CMPPS1}. The formulation of the Ricci-flat derivative expansion is undertaken in Sec. \ref{sec:Ricci-recon}. Here we discuss the important issue of resumming in a closed form the generic expansion. This requires the investigation of another uncharted territory: the higher-derivative curvature-like Carrollian tensors. The Carrollian geometry on the spatial boundary $\mathscr{S}$ is naturally equipped with a (conformal) Carrollian connection, which comes with various curvature tensors presented in Sec.  \ref{sec:Ricci-flat}. The relevant object for discussing the resummability in the anti-de Sitter case is the Cotton tensor, as reviewed in Sec. \ref{sec:resform}. It turns out that this tensor has well-defined Carrollian descendants, which we determine and exploit. With those, the resummability conditions are well-posed and set the framework for obtaining exact Ricci-flat spacetimes in a closed form from conformal-Carrollian-fluid data.

In order to illustrate our results, we provide examples starting from Sec.  \ref{sec:Ricci-flat} and pursuing systematically in Sec.  \ref{sec:ex}. Generic Carrollian perfect fluids are meticulously studied and shown to be dual  to the general Ricci-flat  Kerr--Taub--NUT family. The non perfect Carrollian fluid called Robinson--Trautman fluid is discussed both as the limiting Robinson--Trautman relativistic fluid (Sec. \ref{sec:Ricci-flat}), and alternatively from Carrollian first  principles (Sec. \ref{sec:ex}, following \cite{CMPPS1}). It is shown to be dual to the Ricci-flat Robinson--Trautman spacetime, of which the line element is obtained thanks to our flat resummation procedure. 

One of the resummability requirements is the absence of shear for the Carrollian fluid. This is a geometric quantity, which, if absent, makes possible for using holomorphic coordinates. In App. \ref{holo}, we gather the relevant formulas in this class of coordinates. 

\section{Fluid/gravity in asymptotically locally AdS spacetimes}\label{sec:resform}

We present here an executive summary of the holographic reconstruction of four-dimensional asymptotically locally anti-de Sitter spacetimes from three-dimensional relativistic boundary fluid dynamics. The tool we use is the fluid-velocity derivative expansion. We show that exact Einstein spacetimes written in a closed form can arise by resumming this expansion. It appears that the  key conditions allowing for such an explicit resummation are the absence of shear in the fluid flow, as well as the relationship among the non-perfect components of the fluid energy--momentum tensor (\emph{i.e.} the heat current and the viscous stress tensor) and the boundary Cotton tensor.

\subsection{The derivative expansion}

\subsubsection*{The spirit}

Due to the Fefferman--Graham ambient metric construction \cite{Anderson:2004yi}, asymptotically locally anti-de Sitter four-dimensional spacetimes are determined by a set of independent boundary data, namely  a  three-dimensional  metric  
$\text{d}s^2=g_{\mu\nu}\text{d}x^\mu\text{d}x^\nu$ and a rank-2 tensor
$\text{T}=T_{\mu \nu}\text{d}x^\mu\text{d}x^\nu$,
 symmetric ($T_{\mu\nu}=T_{\nu\mu}$), traceless ($T^\mu_{\hphantom{\mu}\mu}=0$) and conserved:
\begin{equation}
\label{T-cons}
\nabla^\mu T_{\mu\nu}=0.
    \end{equation}
    
Perhaps the most well known subclass of asymptotically locally AdS spacetimes are those whose boundary metrics are conformally flat (see \emph{e.g.} \cite{Skender, Marolf}). These are asymptotically \emph{globally} anti-de Sitter. The asymptotic symmetries of such spacetimes comprise the finite dimensional conformal group, \emph{i.e.} $SO(3,2)$ in four dimensions \cite{Ashtekar}, and AdS/CFT is at work giving rise to a boundary conformal field theory. Then, the rank-2 tensor $T_{\mu\nu}$ is interpreted as the expectation value over a boundary quantum state of the conformal-field-theory energy--momentum tensor. Whenever hydrodynamic regime is applicable, this approach gives  rise to the so-called fluid/gravity correspondence and all its important spinoffs (see \cite{Hubeny:2011hd} for a review).  

For a long time, all the work on fluid/gravity correspondence was confined to asymptotically globally AdS spacetimes, hence to holographic boundary fluids that flow on conformally flat backgrounds. In a series of works \cite{Caldarelli:2012cm, Mukhopadhyay:2013gja, Petropoulos:2014yaa, Gath:2015nxa, Petropoulos:2015fba, Petkou:2015fvh} we have extended the fluid/gravity correspondence into the realm of asymptotically locally AdS$_4$ spacetimes.  In the following, we present and summarize our salient findings.

\subsubsection*{The energy--momentum tensor}

Given the energy--momentum tensor of the boundary fluid and assuming that it represents a state in a hydrodynamic regime, one should be able to pick a boundary congruence $\text{u}$, playing the r\^ole of fluid velocity. Normalizing the latter as\footnote{ This unconventional normalization ensures that the derivative expansion is well-behaved in the $k\to 0$ limit. In the language of fluids, it naturally incorporates the scaling introduced in \cite{Barnich:2012aw} -- see footnote \ref{glenn}.}  $\| \text{u} \|^2=-k^2$ we can in general decompose the energy--momentum tensor as
\begin{equation}\label{T} 
T_{\mu \nu}=(\varepsilon+p) \frac{u_\mu  u_\nu}{k^2} +p  g_{\mu\nu}+   \tau_{\mu \nu}+ \frac{u_\mu  q_\nu}{k^2}+ \frac{u_\nu  q_\mu}{k^2} .
\end{equation}
We assume local thermodynamic equilibrium with $p$ the local pressure
and  $\varepsilon$ the local energy density:  
\begin{equation}
\label{long} 
\varepsilon=\frac{1}{k^2}T_{\mu \nu} u^\mu u^\nu.
\end{equation}
 A local-equilibrium thermodynamic equation of state $p=p(T)$ is also needed for completing the system, and we omit the chemical potential as no independent conserved current, \emph{i.e.} no gauge field in the bulk, is considered here.

The symmetric viscous stress tensor $\tau_{\mu \nu}$ and the heat current $q_\mu$ are purely transverse:
\begin{equation}\label{trans}
u^\mu    \tau_{\mu \nu}=0, \quad u^\mu  q_\mu =0, \quad 
q_\nu= -{\varepsilon} u_\nu-u^\mu  T_{\mu \nu}.
\end{equation}
For a conformal fluid in $3$ dimensions
\begin{equation}\label{confluid}
\varepsilon=2p,\quad  \tau^\mu_{\hphantom{\mu} \mu}=0.
\end{equation}
The quantities at hand  are usually expressed as expansions in temperature and velocity derivatives, the coefficients of which characterize the transport phenomena occurring in the fluid. In first-order hydrodynamics 
\begin{eqnarray}\label{e1}
&&\tau_{(1)\mu \nu}=-2\eta \sigma_{\mu \nu}-\zeta h_{\mu\nu}\Theta,\\
\label{q1} &&q_{(1)\mu}= -\kappa h_\mu^{\hphantom{\mu}\nu}\left(\partial_\nu T+\frac{T}{k^2}\, a_\nu \right),
\end{eqnarray}
where 
$h_{\mu \nu} $ is the projector onto the space transverse to the velocity field:
\begin{equation}
\label{relproj}
h_{\mu\nu}=\dfrac{u_{\mu}u_{\nu}}{k^2}+g_{\mu\nu},
\end{equation}
and\footnote{Our conventions for (anti-) symmetrization are:
$A_{(\mu\nu)}=\frac{1}{2}\left(A_{\mu\nu}+A_{\nu\mu}\right)$ and $ 
A_{[\mu\nu]}=\frac{1}{2}\left(A_{\mu\nu}-A_{\nu\mu}\right)$.}
\begin{eqnarray}
&a_\mu =u^\nu \nabla_\nu u_\mu , \quad
\Theta=\nabla_\mu  u^\mu ,& \label{def21}\\
&\sigma_{\mu \nu}= \nabla_{(\mu } u_{\nu )} + \frac{1}{k^2}u_{(\mu } a_{\nu )} -\frac{1}{2} \Theta h_{\mu \nu}  ,&
\label{def23}
\\ &\omega_{\mu \nu}= \nabla_{[\mu } u_{\nu ]} +  \frac{1}{k^2}u_{[\mu }a_{\nu] },&\label{def24}
\end{eqnarray}
are the acceleration (transverse), the expansion, the shear and the vorticity (both rank-two tensors are transverse and traceless). As usual, $\eta,\zeta$ are the shear and bulk viscosities, and $\kappa$ is the thermal conductivity. 

It is customary to introduce the
vorticity two-form 
\begin{equation}
\label{def25}
\omega=\frac{1}{2}\omega_{\mu\nu }\, \mathrm{d}\mathrm{x}^\mu\wedge\mathrm{d}\mathrm{x}^\nu  
=\frac{1}{2}\left(\mathrm{d}\mathrm{u} +\frac{1}{k^2}
\mathrm{u} \wedge\mathrm{a} \right),
\end{equation}
as well as the Hodge--Poincar\'e dual of this form, which is proportional to  $\text{u}$ (we are in $2+1$ dimensions):
\begin{equation}
\label{omdual}
k\gamma \text{u}=\star\omega\quad\Leftrightarrow\quad k\gamma u_\mu=\frac{1}{2}\eta_{\mu\nu\sigma}\omega^{\nu\sigma},
\end{equation}
where $\eta_{\mu\nu\sigma}=\sqrt{-g} \epsilon_{\mu\nu\sigma}$.
In this expression  $\gamma$ is a scalar, that can also be expressed as
\begin{equation}\label{gamma}
\gamma^2 = \frac{1}{2k^4} \omega_{\mu\nu} \omega^{\mu\nu} .
\end{equation} 

In three spacetime dimensions and in the presence of a vector field, one naturally defines a fully antisymmetric two-index tensor as
\begin{equation}
\label{eta2}
\eta_{\mu\nu}=-\frac{u^\rho}{k}\eta_{\rho\mu\nu}, 
\end{equation} 
obeying
\begin{equation}
\label{eta2contr}
\eta^{\vphantom{\nu}}_{\mu\sigma}\eta_{\nu}^{\hphantom{\nu}\sigma}=h_{\mu\nu}.
\end{equation} 
With this tensor the vorticity reads:
\begin{equation}
\label{eta2vort}
\omega_{\mu\nu}=k^2\gamma \eta_{\mu\nu}. 
\end{equation} 

\subsubsection*{Weyl covariance, Weyl connection and the Cotton tensor}

In the case when the boundary metric $g_{\mu\nu}$ is conformally flat, it was shown that using the above set of boundary data it is possible to reconstruct the  four-dimensional bulk Einstein spacetime order by order in derivatives of the velocity field \cite{Bhattacharyya:2007, Hubeny:2011hd,Haack:2008cp, Bhattacharyya:2008jc}.
The guideline for the spacetime reconstruction based on the derivative expansion is \emph{Weyl covariance}: the bulk geometry should be insensitive to a conformal rescaling of the boundary metric  (weight $-2$)
\begin{equation}
\label{conmet}
\text{d}s^2\to  \frac{\text{d}s^2}{{\cal B}^2},
\end{equation}
which should correspond to a bulk diffeomorphism and be reabsorbed into a redefinition of the radial coordinate:  $r\to{\cal B}\, r$.
At the same time, $u_\mu$ is traded for $\nicefrac{u_\mu}{{\cal B}}$ (velocity one-form), $\omega_{\mu\nu}$ for $ \nicefrac{\omega_{\mu\nu}}{{\cal B}}$ (vorticity two-form) and $T_{\mu\nu}$ for $\mathcal{B} T_{\mu\nu}$. As a consequence, the pressure and energy density have weight $3$, the heat-current $q_\mu$ weight $2$, and the viscous stress tensor $\tau_{\mu\nu}$ weight $1$. 

Covariantization with respect to rescaling requires to introduce a Weyl connection one-form:\footnote{The explicit form of $\text{A}$ is obtained by demanding $\mathscr{D}_{\mu}u^{\mu}=0$ and $u^{\lambda}\mathscr{D}_{\lambda}u_{\mu}=0
$.}
\begin{equation}
\label{Wconc}
\text{A}=\frac{1}{k^2}\left(\text{a} -\frac{\Theta}{2} \text{u}\right),
\end{equation}
which transforms as $\text{A}\to\text{A}-\text{d}\ln {\cal B}$. Ordinary covariant derivatives $\nabla$ are thus traded for Weyl covariant ones $\mathscr{D}=\nabla+w\,\text{A}$, $w$ being the conformal weight of the tensor under consideration.  
We provide for concreteness the Weyl covariant derivative of a weight-$w$ form $v_\mu$:
\begin{equation}
\label{Wv}
\mathscr{D}_\nu v_\mu=\nabla_\nu v_\mu+(w+1)A_\nu v_\mu + A_\mu v_\nu-g_{\mu\nu} A^\rho v_\rho. 
\end{equation}
The Weyl covariant derivative is metric with effective torsion:
\begin{eqnarray}
\mathscr{D}_\rho g_{\mu\nu}&=&0,\\
\left(\mathscr{D}_\mu\mathscr{D}_\nu -\mathscr{D}_\nu\mathscr{D}_\mu\right) f&=& w f F_{\mu\nu},
\end{eqnarray}
where  
\begin{equation}
\label{F}
F_{\mu\nu}=\partial_\mu A_\nu-\partial_\nu A_\mu\end{equation}
is Weyl-invariant.

Commuting the Weyl-covariant derivatives acting on vectors, as usual one defines the Weyl covariant Riemann tensor 
\begin{equation}
\left(\mathscr{D}_\mu\mathscr{D}_\nu -\mathscr{D}_\nu\mathscr{D}_\mu\right) V^\rho=
\mathscr{R}^\rho_{\hphantom{\rho}\sigma\mu\nu} V^\sigma+ w V^\rho F_{\mu\nu}
\end{equation}
($V^\rho$ are weight-$w$)
and the usual subsequent quantities. 
In three spacetime dimensions, the covariant Ricci (weight $0$) and the scalar (weight $2$) curvatures read: 
\begin{eqnarray}
\mathscr{R}_{\mu\nu}&=&{R}_{\mu\nu} + \nabla_\nu A_\mu + A_\mu A_\nu +
g_{\mu\nu}\left(\nabla_\lambda A^\lambda-A_\lambda A^\lambda\right)
-F_{\mu\nu},
\label{curlRic}
\\
\mathscr{R}&=&R +4\nabla_\mu A^\mu- 2 A_\mu A^\mu . \label{curlRc}
\end{eqnarray}
The Weyl-invariant Schouten tensor\footnote{The ordinary Schouten tensor in three spacetime dimensions is given by
$R_{\mu\nu}-\frac{1}{4} R g_{\mu\nu}$.} is 
\begin{equation}
\mathscr{S}_{\mu\nu} = \mathscr{R}_{\mu\nu}-\frac{1}{4} \mathscr{R} g_{\mu\nu}
=S_{\mu\nu}+\nabla_\nu A_\mu + A_\mu A_\nu -\frac{1}{2}A_\lambda A^\lambda g_{\mu\nu}-F_{\mu\nu}.
\end{equation}
Other Weyl-covariant velocity-related 
quantities are 
\begin{eqnarray}
\mathscr{D}_\mu u_{\nu}&=&\nabla_{\mu}u_{\nu}+\frac{1}{k^2}u_{\mu}a_{\nu}-\dfrac{\Theta}{2}h_{\mu\nu}
\nonumber\\&=&\sigma_{\mu\nu}+\omega_{\mu\nu},
\\
\mathscr{D}_\nu\omega^{\nu}_{\hphantom{\nu}\mu}&=&\nabla_\nu\omega^{\nu}_{\hphantom{\nu}\mu},
\\
\mathscr{D}_\nu\eta^{\nu}_{\hphantom{\nu}\mu}&=&2\gamma u_\mu,
\\ u^\lambda
\mathscr{R}_{\lambda\mu}  &=& 
 \mathscr{D}^{\vphantom{(\nu}}_\lambda\left(\sigma^{\lambda}_{\hphantom{\lambda}\mu}-\omega^{\lambda}_{\hphantom{\lambda}\mu}\right)-u^\lambda F_{\lambda\mu} , \label{sigmac}
\end{eqnarray} 
of weights $-1$, $1$, $0$ and  $1$
(the scalar vorticity $\gamma$ has weight $1$). 

The remarkable addition to the fluid/gravity dictionary came with the realization that the derivative expansion can be used to reconstruct Einstein metrics which are asymptotically locally AdS. For the latter, the boundary metric has a non zero Cotton tensor \cite{Caldarelli:2012cm, Mukhopadhyay:2013gja, Petropoulos:2014yaa, Gath:2015nxa, Petropoulos:2015fba}.  The Cotton tensor is generically a three-index tensor with mixed symmetries. In three dimensions, which is the case for our boundary geometry, the Cotton tensor can be dualized into a two-index, symmetric and traceless tensor. It is defined as 
\begin{equation}
C_{\mu\nu}=\eta_{\mu}^{\hphantom{\mu}\rho\sigma}
\mathscr{D}_\rho \left(\mathscr{S}_{\nu\sigma}+F_{\nu\sigma} \right)
=\eta_{\mu}^{\hphantom{\mu}\rho\sigma}
\nabla_\rho \left(R_{\nu\sigma}-\dfrac{R}{4}g_{\nu\sigma} \right)\,.
\label{cotdef}
\end{equation}
The Cotton tensor is Weyl-covariant of weight  $1$ (\emph{i.e.} transforms as $C_{\mu\nu}\to {\cal B}\, C_{\mu\nu}$), and is \emph{identically} conserved: 
\begin{equation}
\label{C-cons}
\mathscr{D}_\rho C^\rho_{\hphantom{\rho}\nu}
=
\nabla_\rho C^\rho_{\hphantom{\rho}\nu}=0,
\end{equation}
sharing thereby all properties of the energy--momentum tensor.    
Following \eqref{T} we can decompose the Cotton tensor into longitudinal, transverse and mixed components with respect to the fluid velocity $\text{u}$:\footnote{Notice that the energy--momentum tensor has an extra factor of $k$ with respect to the Cotton tensor, see \eqref{eqn:Tref}, due to their different dimensions. \label{k-fac}}
\begin{equation}\label{C} 
C_{\mu \nu}=\frac{3c}{2} \frac{u_\mu  u_\nu}{k} + \frac{c k}{2} g_{\mu\nu}-  \frac{c_{\mu \nu}}{k}+ \frac{u_\mu  c_\nu}{k}   +\frac{u_\nu  c_\mu}{k}.
\end{equation}
Such a decomposition naturally defines the weight-$3$ \emph{Cotton scalar density}
\begin{equation}
\label{cottdens} 
c=\frac{1}{k^3}C_{\mu\nu}u^\mu u^\nu,
\end{equation}
as the longitudinal component. The symmetric and  traceless \emph{Cotton stress tensor} $c_{\mu \nu}$  and the \emph{Cotton current} $c_\mu$  (weights 1 and  2, respectively) are purely transverse:
\begin{equation}
\label{cotrans}
  c_{\mu}^{\hphantom{\mu} \mu}=0, \quad
u^\mu   c_{\mu \nu}=0,\quad u^\mu  c_\mu=0,
\end{equation}
and obey
\begin{equation}
\label{cotransp}
c_{\mu\nu}=-k h^{\rho}_{\hphantom{\rho}\mu}h^{\sigma}_{\hphantom{\sigma}\nu}C_{\rho\sigma}+\dfrac{ck^2}{2}h_{\mu\nu}
, \quad 
c_\nu= -c u_\nu-\frac{u^\mu  C_{\mu \nu}}{k}.
\end{equation}

One can use the definition \eqref{cotdef} to further express the Cotton density, current and stress tensor as ordinary or Weyl derivatives of the curvature. We find
\begin{eqnarray}
\label{cotdens}
 c&=&\frac{1}{k^2}u^\nu  \eta^{\sigma\rho} \mathscr{D}_\rho \left(\mathscr{S}_{\nu\sigma}+F_{\nu\sigma} \right),\\
\label{cotcur}
c_\nu&=&\eta^{\rho\sigma}\mathscr{D}_\rho \left(\mathscr{S}_{\nu\sigma}+F_{\nu\sigma} \right)-c u_{\nu} ,\\
c_{\mu\nu} &=& -h^{\lambda}_{\hphantom{\alpha}\mu}\left( k \eta_{\nu}^{\hphantom{\nu}\rho\sigma}-u_{\nu}\eta^{\rho\sigma}\right)\mathscr{D}_\rho \left(\mathscr{S}_{\lambda\sigma}+F_{\lambda\sigma}\right)+\dfrac{c k^2}{2}h_{\mu\nu}.
\label{cotvis}
\end{eqnarray}

\subsubsection*{The bulk Einstein derivative expansion}

Given the ingredients above, the leading terms in a $\nicefrac{1}{r}$ expansion for a four-dimensional Einstein metric are of the form:\footnote{We have traded here the usual advanced-time coordinate used
 in the quoted literature on fluid/gravity correspondence for the retarded time, spelled $t$ 
(see \eqref{ut}).\label{rminusr}}
\begin{eqnarray}
\text{d}s^2_{\text{bulk}} &=&
2\frac{\text{u}}{k^2}(\text{d}r+r \text{A})+r^2\text{d}s^2+\frac{\text{S}}{k^4}\nonumber\\
&&+ \frac{\text{u}^2}{k^4r^2}\left(1 -\frac{1}{2k^4r^2 } \omega_{\alpha\beta} \omega^{\alpha\beta}\right)   \left(\frac{8\pi G T_{\lambda \mu}u^\lambda u^\mu}{k^2 }r+
\frac{C_{\lambda \mu}u^\lambda \eta^{\mu\nu\sigma}\omega_{\nu\sigma}}{2k^4}\right)
\nonumber\\
&&+\text{ terms with $\sigma$, $\sigma^2$, $\nabla \sigma$, \dots}+ \text{O}\left(\mathscr{D}^4\text{u}\right).
\label{papaefgenresc}
\end{eqnarray}
In this expression
\begin{itemize}
\item 
$\text{S}$ is a Weyl-invariant tensor:
\begin{equation}
\label{S}
\text{S}=S_{\mu\nu} \text{d}x^\mu\text{d}x^\nu=-2\text{u}\mathscr{D}_\nu \omega^\nu_{\hphantom{\nu}\mu}\text{d}x^\mu- \omega_\mu^{\hphantom{\mu}\lambda} \omega^{\vphantom{\lambda}}_{\lambda\nu}\text{d}x^\mu\text{d}x^\nu-\text{u}^2\frac{\mathscr{R}}{2} ;
\end{equation}
\item the boundary metric is parametrized \emph{\`a la} Randers--Papapetrou:
\begin{equation}
\label{carrp}
\text{d}s^2 =- k^2\left(\Omega \text{d}t-b_i \text{d}x^i
\right)^2+a_{ij} \text{d}x^i \text{d}x^j;
\end{equation}
\item the boundary conformal fluid velocity field  and the corresponding one form are
\begin{equation}
\label{ut}
\text{u}=\frac{1}{\Omega}\partial_t \quad \Leftrightarrow
\quad\text{u}= -k^2\left(\Omega\text{d}t-b_i\text{d}x^i\right),
\end{equation}
\emph{i.e.}
the fluid is at rest in the frame associated with the coordinates in \eqref{carrp} --  this is not a limitation, as one can always choose a local frame where the fluid is at rest, in which the metric reads  \eqref{carrp}  (with $\Omega$, $b_i$ and $a_{ij}$ functions of all coordinates);
\item $\omega_{\mu\nu}$ is the vorticity of $\text{u}$ as given in \eqref{def24}, which reads:
\begin{equation}
\omega=\dfrac{1}{2}\omega_{\mu\nu}\text{d}x^\mu\wedge\text{d}x^\nu=\frac{k^2}{2}\left(
\partial_{i}b_{j}+\frac{1}{\Omega}b_{i}\partial_{j}\Omega+\frac{1}{\Omega}b_{i}\partial_tb_{j}
\right)\text{d}x^i\wedge \text{d}x^j; \label{omu}
\end{equation}
\item  $\gamma^2= \frac{1}{2}a^{ik}a^{jl}\left(\partial_{[i}b_{j]}+\frac{1}{\Omega}b_{[i}\partial_{j]}\Omega+\frac{1}{\Omega}b_{[i}\partial_tb_{j]}\right)\left(\partial_{[k}b_{l]}+\frac{1}{\Omega}b_{[k}\partial_{l]}\Omega+\frac{1}{\Omega}b_{[k}\partial_tb_{l]}\right)
$;
\item the expansion and acceleration are 
\begin{eqnarray}
\label{carexp} 
\Theta&=&
\dfrac{1}{\Omega}
\partial_t \ln\sqrt{a}
,\\
\text{a}&=&k^2\left(\partial_{i}\ln \Omega+\frac{1}{\Omega}\partial_tb_{i}
\right)\text{d}x^i,
\label{caracclim} 
\end{eqnarray}
leading to the Weyl connection
\begin{equation}
\label{A}
\text{A}=\frac{1}{\Omega} \left(\partial_{i} \Omega+\partial_tb_{i}-
\frac{1}{2}b_{i} \partial_t \ln\sqrt{a}
\right)\text{d}x^i+\frac{1}{2} \partial_t \ln\sqrt{a}\text{d}t\,,
\end{equation}
with $a$ the determinant of $a_{ij}$;
\item $\tfrac{1}{k^2}T_{\mu \nu} u^\mu u^\nu$ is the energy density $\varepsilon$ of the fluid  (see \eqref{long}), and in the Randers--Papapetrou frame associated with \eqref{carrp}, \eqref{ut}, $q_0$, $\tau_{00}$ , $\tau_{0i}=\tau_{i0}$ entering in \eqref{T} all vanish due to \eqref{trans};
\item $\frac{1}{2k^4}C_{\lambda \mu}u^\lambda \eta^{\mu\nu\sigma}\omega_{\nu\sigma}=c\gamma$, 
where we have used \eqref{omdual} and \eqref{cottdens}, and similarly $c_0=c_{00}=c_{0i}=c_{i0}=0$ as a consequence of  \eqref{cotrans}  with \eqref{carrp}, \eqref{ut};
\item $\sigma$, $\sigma^2$, $\nabla \sigma$ stand for the shear of $\text{u}$ and combinations of it, as computed from \eqref{def23}:
\begin{equation}
\label{sig}
\sigma=\frac{1}{2\Omega}\left(\partial_t a_{ij}-a_{ij}\partial_t \ln \sqrt{a}\right)\text{d}x^i\text{d}x^j.
\end{equation}
\end{itemize}
We have not exhibited explicitly shear-related terms because we will ultimately assume the absence of shear for our congruence. This raises the important issue of choosing the fluid velocity field, not necessary in the Fefferman--Graham expansion, but fundamental here. In relativistic fluids, the absence of sharp distinction between heat and matter fluxes leaves a freedom in setting the velocity field. This choice of \emph{hydrodynamic frame} is not completely arbitrary though, and one should stress some reservations, which are 
often dismissed, in particular in the already quoted fluid/gravity literature. 

As was originally exposed in \cite{Landau} and extensively discussed \emph{e.g.} in \cite{Kovtun:2012rj}, the fluid-velocity ambiguity is well posed in the presence of a conserved current $\text{J}$, 
naturally decomposed into a longitudinal perfect piece and a transverse part: 
\begin{equation}
\label{curdec}
J^\mu=\varrho u^\mu +j^\mu.
\end{equation}
The velocity freedom originates from the redundancy in the heat current $\text{q}$ and the non-perfect piece of the matter current $\text{j}$. One may therefore set $\text{j}=0$ and reach the Eckart frame. Alternatively $\text{q}=0$ defines the Landau--Lifshitz frame. In the absence of matter current, nothing guarantees that one can still move to the Landau--Lifshitz frame, and setting  $\text{q}=0$ appears as a constraint on the fluid, rather than a choice of frame for describing arbitrary fluids. This important issue was recently discussed in the framework of holography \cite{Ciambelli:2017wou}, from which it is clear that setting $\text{q}=0$ in the absence of a conserved current would simply inhibit certain classes of Einstein spaces to emerge holographically from boundary data, and possibly blur
the physical phenomena occurring in the fluids under consideration. Consequently, we will not make any such assumption, keeping the heat current as part of the physical data. 

We would like to close this section with an important comment on asymptotics. The reconstructed bulk spacetime can be asymptotically locally or globally anti-de Sitter. This property is read off directly inside terms appearing at designated orders in the radial expansion, and built over specific boundary tensors. 
For $d+1$-dimensional boundaries, the boundary energy--momentum contribution first appears at order $\nicefrac{1}{r^{d-1}}$, whereas the boundary  Cotton tensor\footnote{ Actually, the object appearing in generic dimension is the Weyl divergence of the boundary Weyl tensor, which contains also the Cotton tensor (see  \cite{MH}\label{marie} for a preliminary discussion on  this point).} emerges at order $\nicefrac{1}{r^{2}}$. This behaviour is rooted in the Eddington--Finkelstein gauge used in \eqref{papaefgenresc}, but appears also in the slightly different  Bondi gauge.
It is however absent in the Fefferman--Graham coordinates, where the Cotton cannot be possibly isolated in the expansion.

\subsection{The resummation of AdS spacetimes}\label{resAdS}

\subsubsection*{Resummation and exact Einstein spacetimes in closed form}

In order to further probe the derivative expansion \eqref{papaefgenresc}, we will impose the fluid velocity congruence be shearless. This choice has the virtue of reducing considerably the number of terms compatible with conformal invariance in  \eqref{papaefgenresc}, and potentially making this expansion resummable, thus leading to an Einstein metric written in a closed form. Nevertheless, this  shearless condition  reduces the class of Einstein spacetimes that can be reconstructed holographically to the algebraically special ones \cite{Mukhopadhyay:2013gja, Petropoulos:2014yaa,Gath:2015nxa,Petropoulos:2015fba,Petkou:2015fvh}. Going beyond this class is an open problem that we will not address here.

Following \cite{Bhattacharyya:2008jc, Mukhopadhyay:2013gja, Petropoulos:2014yaa,Gath:2015nxa,Petropoulos:2015fba,Petkou:2015fvh}, it is tempting to try a resummation  of  \eqref{papaefgenresc} using the following substitution:
\begin{equation}
\label{resubst}
1 -\frac{\gamma^2}{r^2 } \to \frac{r^2}{\rho^2}
\end{equation}
with
\begin{equation}\label{rho2c}
 \rho^2= r^2 +\gamma^2.
\end{equation} 
The resummed expansion would then read
\begin{equation}
\boxed{
\text{d}s^2_{\text{res. Einstein}} =
2\frac{\text{u}}{k^2}(\text{d}r+r \text{A})+r^2\text{d}s^2+\frac{\text{S}}{k^4}
+ \frac{\text{u}^2}{k^4\rho^2} \left(8\pi G \varepsilon r+
c \gamma\right),}
\label{papaefgenrescrec}
\end{equation}
which is indeed written in a closed form. 
Under the conditions listed below, the metric  \eqref{papaefgenrescrec} defines the line element of an \emph{exact} Einstein space with $\Lambda=-3 k^2$.
\begin{itemize}
\item \textsl{The congruence $\text{u}$ is shearless.} This requires (see \eqref{sig})
\begin{equation}
\label{shear3d}
\partial_t a_{ij}=a_{ij}\partial_t \ln \sqrt{a}.
\end{equation}
Actually \eqref{shear3d} is equivalent to ask that the two-dimensional spatial section $\mathscr{S}$ defined at every time $t$ and equipped with the metric $\text{d}\ell^2=a_{ij} \text{d}x^i \text{d}x^j$ 
is conformally flat. This may come as a surprise because every two-dimensional metric is conformally flat. However, $a_{ij} $ generally depends on space $\textbf{x}$ \emph{and} time $t$, and the transformation required to bring it in a form proportional to the flat-space metric might depend on time. This would spoil the three-dimensional structure \eqref{carrp} and alter the \emph{a priori} given $\text{u}$. Hence, $\text{d}\ell^2$ is conformally flat within the three-dimensional spacetime \eqref{carrp} under the condition that the transformation used to reach the explicit conformally flat form be of the type $ \textbf{x}^{\prime}=\textbf{x}^{\prime}(\textbf{x})$. This exists if and only if  \eqref{shear3d} is satisfied.\footnote{A peculiar subclass where this works is when $\partial_t$ is a Killing field.} Under this condition,  one can always choose $\zeta=\zeta(\mathbf{x})$, $\bar\zeta=\bar\zeta(\mathbf{x})$ such that
\begin{equation}
\label{adfr2-CF}
\text{d}\ell^2=a_{ij}\,\mathrm{d}x^i \mathrm{d}x^j=\frac{2}{P^2}\text{d}\zeta\text{d}\bar\zeta
\end{equation}
with $P=P(t,\zeta,\bar \zeta)$ a real function.
 Even though this does not hold for arbitrary $\text{u}=\nicefrac{\partial_t }{\Omega}$, one can show that there exists always a congruence for which it does \cite{Coll}, and this will be chosen for the rest of the paper. 
\item \textsl{The heat current of the boundary fluid introduced in \eqref{T} and \eqref{trans} is identified with the transverse-dual of the Cotton current defined in  \eqref{C} and \eqref{cotransp}.} The Cotton current being transverse to $\text{u}$, it defines a field on the conformally flat two-surface $\mathscr{S}$, the existence of which is guaranteed by the absence of shear. This surface is endowed with a natural 
hodge duality mapping a vector onto another, which can in turn be lifted back to the three-dimensional spacetime as a new transverse vector. This whole process is taken care of by the action of $\eta^\nu_{\hphantom{\nu}\mu}$ defined in \eqref{eta2}: 
\begin{equation}
\boxed{
q_\mu=\frac{1}{8\pi G}\eta^\nu_{\hphantom{\nu}\mu}c_\nu\\
=
\frac{1}{8\pi G}\eta^\nu_{\hphantom{\nu}\mu}\eta^{\rho\sigma}
 \mathscr{D}_\rho
\left(\mathscr{S}_{\nu\sigma}+F_{\nu\sigma} \right),
}
\label{heat-resum}
\end{equation}
where we used \eqref{cotcur} in the last expression. Using holomorphic and antiholomorphic coordinates $\zeta,\bar\zeta$ as in \eqref{adfr2-CF}\footnote{Orientation is chosen such that in the coordinate frame $\eta_{0\zeta\bar\zeta}=\sqrt{-g}\epsilon_{0\zeta\bar\zeta}=\frac{\text{i}\Omega}{P^2}$, where $x^0=kt$.
\label{orient}}
leads to $\eta^\zeta_{\hphantom{\zeta}\zeta}= \text{i} $ and $\eta^{\bar\zeta}_{\hphantom{\bar\zeta}\bar\zeta}=- \text{i} $, and thus
\begin{equation}
\text{q}=\frac{ \text{i}}{8\pi G} \left( c_\zeta \text{d}\zeta
- c_{\bar\zeta} \text{d}\bar\zeta
\right).
\label{heat-resum-hol}
\end{equation}

\item \textsl{The viscous stress tensor of the boundary conformal fluid introduced in \eqref{T} is identified with the transverse-dual of the Cotton stress tensor defined in  \eqref{C} and \eqref{cotransp}.} Following the same pattern as for the heat current, we obtain:
\begin{equation}
\boxed{
\begin{array}{rcl}
\tau_{\mu\nu}&=&-\frac{1}{8\pi Gk^2}\eta^\rho_{\hphantom{\rho}\mu}c_{\rho\nu}\\
&=&\frac{1}{8\pi Gk^2}\left(-\frac{1}{2}u^\lambda
\eta_{\mu\nu} \eta^{\rho\sigma}+\eta^\lambda_{\hphantom{i}\mu} \left(k\eta_{\nu}^{\hphantom{\nu}\rho\sigma}
-u_\nu\eta^{\rho\sigma}\right)
\right)
 \mathscr{D}_\rho 
\left(\mathscr{S}_{\lambda\sigma}+F_{\lambda\sigma} 
\right),
\end{array}
}
\label{visc-resum}
\end{equation}
where we also used \eqref{cotvis} in the last equality. The viscous stress tensor $\tau_{\mu\nu}$ is transverse symmetric and traceless because these are the properties of the Cotton stress tensor $c_{\mu\nu}$. Similarly, we find in complex coordinates:
\begin{equation}
\tau=-\frac{\text{i} }{8\pi Gk^2} \left(c_{\zeta\zeta} \text{d}\zeta^2
-c_{\bar\zeta\bar\zeta} \text{d}\bar\zeta^2
\right).
\label{visc-resum-hol}
\end{equation}
\item\textsl{The energy--momentum tensor defined in \eqref{T} with $p=\nicefrac{\varepsilon}{2}$, heat current as in \eqref{heat-resum} and viscous stress tensor as in \eqref{visc-resum} must be conserved,} \emph{i.e.}\textsl{ obey Eq. \eqref{T-cons}.} These are differential constraints that from a bulk perspective can be thought of as a generalization of the Gauss law. 
\end{itemize}

Identifying parts of the energy--momentum tensor with the Cotton tensor may be viewed as setting integrability conditions, similar to the electric--magnetic duality conditions in electromagnetism, or in Euclidean gravitational dynamics. As opposed to the latter, it is here implemented in a rather unconventional manner, on the conformal boundary.

It is important to emphasize
that the conservation equations  \eqref{T-cons} concern \emph{all} boundary data. 
On the fluid side the only remaining unknown piece is the energy density $\varepsilon(x)$, whereas for the boundary metric $\Omega(x)$, $b_i(x)$ and $a_{ij}(x)$ are available and must obey \eqref{T-cons},  together with $\varepsilon(x)$. Given these ingredients, \eqref{T-cons} turns out to be precisely the set of equations obtained by demanding bulk Einstein equations be satisfied with the metric \eqref{papaefgenrescrec}. This observation  is at the heart of our analysis. 

\subsubsection*{The bulk algebraic structure and the physics of the boundary fluid}

The pillars of our approach are (\romannumeral1) the requirement of a shearless fluid congruence and (\romannumeral2) the identification of the non-perfect energy--momentum tensor  pieces with the corresponding Cotton components by transverse dualization.

What does motivate these choices? 
The answer to this question is rooted to the Weyl tensor and to the remarkable integrability properties its structure can provide to the system. 

Let us firstly recall that from the bulk perspective, $\text{u}$ is a manifestly null congruence associated with the vector $\partial_r$\,. 
One can show (see \cite{Petropoulos:2015fba}) that this bulk congruence is also \emph{geodesic} and \emph{shear-free}. 
Therefore, accordingly to the generalizations of the Goldberg--Sachs theorem, if the bulk metric 
\eqref{papaefgenresc} is an Einstein space, then it is algebraically special, \emph{i.e.} of Petrov type II, III, D, N or O. Owing to the close relationship between the algebraic structure and the integrability properties of Einstein equations, it is clear why the absence of shear in the fluid congruence plays such an instrumental r\^ole in making the tentatively resummed expression \eqref{papaefgenrescrec} an exact Einstein space.

The structure of the bulk Weyl tensor makes it possible to go deeper in foreseeing how the boundary data should be tuned in order for the resummation to be successful. Indeed the Weyl tensor can be expanded for large-$r$, and the dominant term ($\nicefrac{1}{r^3}$) exhibits the following combination of the boundary energy--momentum and Cotton tensors \cite{Mansi:2008br, Mansi:2008bs, deHaro:2008gp, Ioannis, Olea}:
\begin{equation}
\label{eqn:Tref}
	T_{\mu\nu}^\pm = T_{\mu\nu} \pm \frac{\text{i}}{8\pi G k }C_{\mu\nu},
\end{equation}
satisfying a conservation equation, analogue to \eqref{T-cons}
\begin{equation}
\label{Tref-cons}
\nabla^\mu T^\pm_{\mu\nu}=0.
    \end{equation}

For algebraically special spaces, these complex-conjugate tensors simplify considerably (see detailed discussions in \cite{Mukhopadhyay:2013gja, Petropoulos:2014yaa,Gath:2015nxa,Petropoulos:2015fba,Petkou:2015fvh}), and this suggests the transverse duality enforced between the Cotton and the energy--momentum non-perfect components. Using \eqref{heat-resum-hol} and \eqref{visc-resum-hol}, we find indeed for the tensor $\text{T}^+$ in complex coordinates:
\begin{equation}
\label{eqn:T+}
\text{T}^+=\left(\varepsilon +\frac{\text{i}  c}{8\pi G}\right)\left(
\frac{\text{u}^2}{k^2}+\frac{1}{2}\text{d}\ell^2\right)+\frac{\text{i} }{4\pi G k^2}
\left(2 c_\zeta \text{d}\zeta \text{u}-c_{\zeta\zeta} \text{d}\zeta^2
\right),
\end{equation}
and similarly for $\text{T}^-$ obtained by complex conjugation with
\begin{equation}
\label{eqn:epm}
\varepsilon_\pm=\varepsilon \pm\frac{\text{i}  c}{8\pi G}.
\end{equation}
The bulk Weyl tensor and consequently the Petrov class of the bulk Einstein space are encoded in the three complex functions of the boundary coordinates:
$\varepsilon_+$, $c_\zeta$ and $c_{\zeta\zeta}$.

The proposed resummation procedure, based on boundary relativistic fluid dynamics of non-perfect fluids with heat current and stress tensor designed from the boundary Cotton tensor, allows to reconstruct all algebraically special four-dimensional Einstein spaces. The simplest correspond to a Cotton tensor of the perfect form \cite{Mukhopadhyay:2013gja}. The complete class of  Pleba\'nski--Demia\'nski family  \cite{Plebanski:1976gy} requires non-trivial $b_i$ with two commuting Killing fields \cite{Petropoulos:2015fba}, while vanishing $b_i$ without isometry leads to the Robinson--Trautman Einstein spaces \cite{Gath:2015nxa}. For the latter, the heat current and the stress tensor obtained from the Cotton by the transverse duality read:
\begin{eqnarray}
\label{RT-q}
\text{q}&=& -\frac{1}{16\pi G}\left(\partial_{\zeta}K\text{d}\zeta+\partial_{\bar\zeta}K\text{d}\bar\zeta
\right),
\\
\label{RT-tau}
\tau &=& \frac{1}{8\pi G k^2P^2}
\left(
\partial_{\zeta} \left(P^2\partial_t\partial_{\zeta}
\ln P
\right)\text{d}\zeta^2+ 
\partial_{\bar\zeta} \left(P^2\partial_t\partial_{\bar\zeta}
\ln P
\right)
\text{d}\bar\zeta^2 \right),
\end{eqnarray}
where $K= 2P^2 \partial_{\bar\zeta} \partial_\zeta\ln P$ is the Gaussian curvature of \eqref{adfr2-CF}. 
With these data the conservation of the energy--momentum tensor \eqref{T-cons} enforces the absence of spatial dependence in $\varepsilon=2p$, and 
leads to a single independent equation, the heat equation:
\begin{equation}
\label{RT}
12 M \partial_t \ln P+ \Delta K=4\partial_t M.
\end{equation} 
This is the Robinson--Trautman equation, here expressed in terms of $M(t)=4\pi G \varepsilon(t)$.

The boundary fluids emerging in the systems considered here have a specific physical behaviour. This behaviour is inherited from the boundary geometry, since their excursion away from perfection is encoded in the Cotton tensor via the transverse duality. In the hydrodynamic frame at hand, this implies in particular that the derivative expansion of the energy--momentum tensor terminates at third order.
Discussing this side of the holography is not part of our agenda. We shall only stress that such an analysis does not require to change hydrodynamic frame. Following \cite{Ciambelli:2017wou}, it is possible to show that the frame at hand is the Eckart frame. Trying to discard the heat current in order to reach a Landau--Lifshitz-like frame (as in \cite{deFreitas:2014lia, Bakas:2014kfa, Bakas:2015hdc, Skenderis:2017dnh} for Robinson--Trautman) is questionable, as already mentioned earlier, because of the absence of conserved current, and distorts the physical phenomena occurring in the holographic conformal fluid.

\section{The Ricci-flat limit \RNum{1}: Carrollian geometry and Carrollian fluids}\label{sec:Ricci-flat}

The Ricci-flat limit is achieved at vanishing $k$. Although no conformal boundary exists in this case, 
a two-dimensional spatial conformal structure emerges at null infinity. Since the Einstein bulk spacetime derivative expansion is performed along null tubes, it provides the appropriate arena for studying both the nature of the  two-dimensional ``boundary'' and the dynamics of the degrees of freedom it hosts as ``holographic duals'' to the bulk Ricci-flat spacetime. 

\subsection{The Carrollian boundary geometry}\label{sec:cargeo}

\subsubsection*{The emergence of a boundary}

{  For vanishing $k$, time decouples in the boundary geometry \eqref{carrp}. 
There exist two decoupling limits, associated with two distinct contractions of the Poincar\'e group: the Galilean, reached at infinite velocity of light and referred to as ``non-relativistic'', and the Carrollian, emerging at zero velocity of light \cite{Levy} -- often called ``ultra-relativistic''}. In  \eqref{carrp}, $k$ plays effectively the r\^ole of velocity of light and $k\to0$ is indeed a \emph{Carrollian limit}.

This very elementary observation sets precisely and unambiguously the fate of asymptotically flat 
holography: \emph{the reconstruction of four-dimensional Ricci-flat spacetimes is based on Carrollian boundary geometry}.

{ The appearance of Carrollian symmetry, or better, conformal Carrollian symmetry at null infinity of asymptotically flat spacetimes is not new \cite{Duval:2014uoa,Duval:2014uva, Duv, DuvG}. It has attracted attention in the framework of flat holography, mostly from the algebraic side \cite{Att1,Att2}, or in relation with its dual geometry emerging in the Galilean limit, known as Newton--Cartan  
(see \cite{NC}). The novelties we bring in the present work are twofold.} On the one hand, the Carrollian geometry emerging at null infinity is generally non-flat, \emph{i.e.} it is not isometric under the Carroll group, 
but under a more general group associated with a time-dependent positive-definite spatial metric and a Carrollian time arrow, this general Carrollian geometry being covariant under a subgroup of the diffeomorphisms dubbed Carrollian diffeomorphisms. On the other hand, the Carrollian surface is the natural host for a Carrollian fluid, zero-$k$ limit of the relativistic  boundary fluid dual to the original Einstein space of which we consider the flat limit. This Carrollian fluid must be considered as the holographic dual of a Ricci-flat spacetime, and its dynamics (studied in Sec. \ref{sec:carfluid}) as the dual of gravitational bulk dynamics at zero cosmological constant. { From the hydrodynamical viewpoint, this gives a radically new perspective on the subject of flat holography.}

\subsubsection*{The Carrollian geometry: connection and curvature}

The Carrollian geometry consists of a spatial surface $\mathscr{S}$ endowed with a positive-definite metric
\begin{equation}
\label{dmet}
\text{d}\ell^2=a_{ij} \text{d}x^i \text{d}x^j,
\end{equation}
and a Carrollian time $t\in \mathbb{R}$.\footnote{We are genuinely describing a spacetime 
$\mathbb{R}\times \mathscr{S}$ endowed with a Carrollian structure, and this is actually how the boundary geometry should be spelled. In order to make the distinction with the relativistic pseudo-Riemannian  three-dimensional spacetime boundary $\mathscr{I}$ of AdS bulks, we quote only the spatial surface $\mathscr{S}$ when referring to the Carrollian boundary geometry of a Ricci-flat bulk spacetime. For a complete description of such geometries we recommend \cite{Bekaert:2015xua}.} The metric on $\mathscr{S}$ is generically time-dependent:
$a_{ij} =a_{ij} (t,\mathbf{x})$.
Much like a Galilean space is observed from a spatial frame moving with respect to a local inertial frame with velocity $\mathbf{w}$, a Carrollian frame is described by a form $\pmb{b}=b_i(t,\mathbf{x})\, \text{d}x^i$.  The latter is \emph{not} a velocity because in Carrollian spacetimes motion is forbidden. It is rather an inverse velocity, describing a ``temporal frame'' and plays a dual r\^ole. A scalar $\Omega(t,\mathbf{x})$ is also introduced (as in the Galilean case, see \cite{CMPPS1} -- this reference will be useful along the present section), as it may naturally arise from the $k\to 0$ limit. 

We define the Carrollian diffeomorphisms as
\begin{equation}
\label{cardifs} 
t'=t'(t,\textbf{x})\quad \text{and} \quad \textbf{x}^{\prime}=\textbf{x}^{\prime}(\textbf{x})
\end{equation}
with Jacobian functions  
\begin{equation}
\label{carj}
J(t,\mathbf{x})=\frac{\partial t'}{\partial t},\quad j_i(t,\mathbf{x}) = \frac{\partial  t'}{\partial x^{i}},\quad 
J^i_j(\mathbf{x}) = \frac{\partial x^{i\prime}}{\partial x^{j}}.
\end{equation}
Those are the diffeomorphisms adapted to the Carrollian geometry since under such transformations, d$\ell^2$ remains a positive-definite metric (it does not produce terms involving $\text{d}t'$). Indeed, 
\begin{equation}
\label{cardifabom}
a^{\prime}_{ij} =a_{kl} J^{-1k}_{\hphantom{-1}i} J^{-1l}_{\hphantom{-1}j} ,\quad
b^{\prime}_{k}=\left( b_i+\frac{\Omega}{J} j_i\right)J^{-1i}_{\hphantom{-1}k},\quad
\Omega^{\prime }=\frac{\Omega}{J},\quad
\end{equation}
whereas the time and space derivatives become
\begin{equation}
\label{cartjj}
\partial^\prime_t=\frac{1}{J}\partial_t,\quad
\partial^\prime_j=J^{-1i}_{\hphantom{-1}j}\left(\partial_i-
\frac{j_i}{J}\partial_t\right).
\end{equation}
We will show in a short while that the Carrollian fluid equations  are precisely covariant under this particular set of diffeomorphisms.

Expression \eqref{cartjj} shows that the ordinary exterior derivative of a scalar function does not transform as a form. To overcome this issue, it is desirable to introduce a Carrollian derivative as 
\begin{equation}
\label{dhat}
\hat\partial_i=\partial_i+\frac{b_i}{\Omega}\partial_t,
\end{equation}
transforming as 
\begin{equation}
\label{carb2}
\hat\partial_i^\prime =
J^{-1j}_{\hphantom{-1}i} \hat\partial_j.
\end{equation}
Acting on scalars this provides a form, whereas for any other tensor it must be covariantized by introducing a new connection for Carrollian geometry, called \emph{Levi--Civita--Carroll} connection, whose coefficients are the \emph{Christoffel--Carroll} symbols,\footnote{\label{dgamma} We remind that the ordinary  Christoffel symbols are
$\gamma^i_{jk}=\dfrac{a^{il}}{2}\left(\partial_j a_{lk}+\partial_k a_{lj}-\partial_l a_{jk}\right)$.}
\begin{equation}
\label{dgammaCar}
\hat\gamma^i_{jk}=\dfrac{a^{il}}{2}\left(
\hat\partial_j
a_{lk}+\hat\partial_k  a_{lj}-
\hat\partial_l a_{jk}\right)
=\gamma^i_{jk}+c^i_{jk}.
\end{equation}

The Levi--Civita--Carroll covariant derivative acts symbolically as 
$\hat{ \pmb{\nabla}}=\hat{\pmb{\partial} }
+\hat{ \pmb{\gamma}}
$. It is metric and torsionless:
$
\hat\nabla_ia_{jk}=0$, $\hat t^k_{\hphantom{k}ij}=2\hat\gamma^k_{[ij]}=0$.
There is however an effective torsion, since the derivatives $\hat\nabla_i$ do not commute, even when acting of scalar functions $\Phi$ -- where they are identical to $\hat\partial_i$ :
\begin{equation}
\label{carcontor}
[\hat\nabla_i,\hat\nabla_j]\Phi=
\frac{2}{\Omega}\varpi_{ij}\partial_t \Phi.
\end{equation}
Here $\varpi_{ij}$ is a two-form identified as the Carrollian vorticity defined using the Carrollian acceleration one-form $\varphi_i$:
\begin{eqnarray}
\label{caracc}
&\varphi_i=\dfrac{1}{\Omega}\left(\partial_t b_i+\partial_i \Omega\right)
=\partial_t \dfrac{b_i}{\Omega}+\hat\partial_i \ln \Omega
, &\\
\label{carom}
&\varpi_{ij}=\partial_{[i}b_{j]}+b_{[i}\varphi_{j]}=
\dfrac{\Omega}{2}\left(
\hat\partial_{i}\dfrac{b_{j}}{\Omega}
-
\hat\partial_{j}\dfrac{b_{i}}{\Omega}
\right) 
.&
\end{eqnarray}
Since the original relativistic fluid is at rest, the kinematical ``inverse-velocity'' variable potentially present in the Carrollian limit vanishes.\footnote{ A Carrollian fluid is always at rest, but could generally be obtained from a relativistic fluid moving at $v^i= k^2 \beta^i+\text{O}\left(k^4\right)$. In this case, the  
``inverse velocity'' $\beta^i$ would contribute to the kinematics and the dynamics of the fluid (see  \cite{CMPPS1}). Here, $v^i=0$ before the limit $k\to 0$ is taken, so $\beta^i=0$.} Hence the various kinematical quantities such as the vorticity and the acceleration are purely geometric and originate from the temporal Carrollian frame used to describe the surface $\mathscr{S}$. As we will see later, they turn out to be $k\to 0$ counterparts of their relativistic homologues  defined in \eqref{def21}, \eqref{def23}, \eqref{def24} (see also \eqref{carshexp-tempcon} for the expansion and shear). 

The time derivative transforms as in  \eqref{cartjj}, and acting on any tensor under Carrollian diffeomorphisms, it provides another tensor.  This ordinary time derivative has nonetheless an unsatisfactory feature: its action on the metric does not vanish.  One is tempted therefore to
set a new time derivative  $\hat \partial_t$ such that
$
\hat \partial_ta_{jk}=0,
$ while keeping the transformation rule under Carrollian diffeomorphisms:
${\hat\partial}^\prime_t=\frac{1}{J}\hat\partial_t$. This is achieved by introducing a ``temporal Carrollian connection''
\begin{equation}
\label{dgammaCartime}
\hat\gamma^i_{\hphantom{i}j}=\frac{1}{2\Omega}a^{ik}\partial_t a_{kj},
\end{equation}
which allows us to define the time covariant derivative on a vector field:
\begin{equation}
\frac{1}{\Omega}\hat \partial_tV^i=\frac{1}{\Omega} \partial_tV^i+\hat\gamma^i_{\hphantom{i}j} V^j ,\label{Cartimecovdervec}
\end{equation}
while on a scalar the action is as the ordinary time derivative: $\hat \partial_t \Phi=\partial_t \Phi$. Leibniz rule allows extending the action of this derivative to any tensor.

Calling $\hat\gamma^i_{\hphantom{i}j}$ a connection is actually misleading because it transforms as a genuine tensor under Carrollian diffeomorphisms:
$
\hat\gamma^{\prime k}_{\hphantom{\prime k}j}=
J^k_n 
J^{-1m}_{\hphantom{-1}j}
\hat\gamma^n_{\hphantom{n}m}$. Its trace and traceless parts have a well-defined kinematical interpretation, as the expansion and shear, completing the acceleration and vorticity introduced earlier in \eqref{caracc},  \eqref{carom}:
\begin{equation}
\label{carshexp-tempcon}
\theta=
\hat\gamma^{i}_{\hphantom{i}i}=
\dfrac{1}{\Omega}  \partial_t \ln\sqrt{a}  ,\quad           
 \xi^{i}_{\hphantom{i}j}=
\hat\gamma^i_{\hphantom{i}j}-\frac{1}{2}\delta^i_j
\theta
=\frac{1}{2\Omega}a^{ik}\left(\partial_t a_{kj}-
a_{kj}\partial_t \ln\sqrt{a}
\right).
\end{equation}

We can define the curvature associated with a connection, by computing the commutator of covariant derivatives acting on a vector field. We find
\begin{equation}
\left[\hat\nabla_k,\hat\nabla_l\right]V^i
= \hat r^i_{\hphantom{i}jkl}V^j+
\varpi_{kl}\frac{2}{\Omega}\partial_{t}V^i,
\label{veccom}
\end{equation}
where  
\begin{equation}
\label{carriemann}
\hat r^i_{\hphantom{i}jkl} = 
\hat\partial_k\hat\gamma^i_{lj}
-\hat\partial_l\hat\gamma^i_{kj}
+\hat\gamma^i_{km}\hat\gamma^m_{lj}
-\hat\gamma^i_{lm}\hat\gamma^m_{kj}
\end{equation}
is a genuine tensor under Carrollian diffeomorphisms, the Riemann--Carroll tensor. 

As usual, the Ricci--Carroll tensor is
\begin{equation}
\label{carricci}
\hat r_{ij}=\hat r^k_{\hphantom{k}ikj}.
\end{equation}
It is \emph{not} symmetric in general ($\hat r_{ij}\neq \hat r_{ji}$) and carries four independent components:
\begin{equation}
\label{carricci-expand}
\hat r_{ij}=\hat s_{ij}+\hat K a_{ij}+\hat A \eta_{ij}.
\end{equation}
In this expression $\hat s_{ij}$ is symmetric and traceless, whereas\footnote{We use 
$\eta_{ij}=\sqrt{a}\epsilon_{ij}$, which matches, in the zero-$k$ limit, with the spatial components of the $\eta_{\mu\nu}$ introduced in \eqref{eta2}. To avoid confusion we also quote 
that $\eta^{il}\eta_{jl}=\delta^i_j$ and  $\eta^{ij}\eta_{ij}=2$.}
\begin{equation}
\label{scalar}
\hat K=\frac{1}{2} a^{ij}\hat r_{ij}=\frac{1}{2} \hat r, \quad \hat A=\frac{1}{2} \eta^{ij}\hat r_{ij}=\ast\varpi\theta
\end{equation}
are the scalar-electric and scalar-magnetic Gauss--Carroll curvatures, with
\begin{equation}
\label{starvarpi}
\ast\varpi=\frac{1}{2} \eta^{ij} \varpi_{ij}.
\end{equation}

Since time and space are intimately related in Carrollian geometry, curvature extends also in time. This can be seen by computing the covariant time and space derivatives commutator: 
\begin{equation}
\left[\frac{1}{\Omega}\hat\partial_{t},\hat\nabla_i\right]V^i= -2\hat r_{i}V^i+
\left(
\theta\delta_i^j-\hat\gamma^{j}_{\hphantom{j}i}\right)\varphi_{j}
V^i+
\left(\varphi_{i}\frac{1}{\Omega}\hat\partial_{t}
-\hat\gamma^{j}_{\hphantom{j}i}\hat\nabla_j
\right)V^i
.
\label{carriemanntimetilde}
\end{equation}
A Carroll curvature one-form emerges thus as 
\begin{equation}
\hat r_{i}=\frac{1}{2}\left(\hat\nabla_j\xi^{j}_{\hphantom{j}i}
 -\frac{1}{2}\hat\partial_i\theta 
\right).
\label{carriemanntime}
\end{equation}

The Ricci--Carroll curvature tensor $\hat r_{ij}$ and the Carroll curvature one-form $\hat r_{i}$ are actually the Carrollian vanishing-$k$ contraction of the ordinary Ricci tensor $R_{\mu\nu}$ associated with the original three-dimensional pseudo-Riemannian
AdS boundary $\mathscr{I}$, of  Randers--Papapetrou type \eqref{carrp}. The identification of the various pieces is however a subtle task because in this kind of limit, where the size of one dimension shrinks,  the curvature usually develops divergences. From the perspective of the final Carrollian geometry this does not produce any harm because the involved components decouple.

The metric \eqref{dmet} of the Carrollian geometry on $\mathscr{S}$ may or may not be recast in conformally flat form  \eqref{adfr2-CF} using Carrollian diffeomorphisms \eqref{cardifs},  \eqref{carj}. 
A necessary and sufficient condition is the vanishing of the Carrollian shear $\xi_{ij}$, displayed in \eqref{carshexp-tempcon}. Assuming this holds, one proves that the traceless and symmetric piece of the Ricci-Carroll tensor is zero,
\begin{equation}
\label{cars-van}
\hat s_{ij}=0.
\end{equation}
We gather in App. \ref{holo} various expressions when holomorphic coordinates are used and the metric is given in conformally flat form. The absence of shear will be imposed again in Sec. \ref{sec:Ricci-recon}, where it plays a crucial r\^ole in the resummation of the derivative expansion.

\subsubsection*{The conformal Carrollian geometry}

In the present set-up, the spatial surface  $\mathscr{S}$ appears as the null infinity of the resulting Ricci-flat geometry \emph{i.e.} as $\mathscr{I}^+$. This is not surprising. 
The bulk congruence tangent to $\partial_r$ is lightlike. Hence the holographic limit $r\to \infty$ is lightlike, already at finite $k$, which is a well known feature of the derivative expansion, expressed by construction in Eddington--Finkelstein-like coordinates \cite{Bhattacharyya:2007, Hubeny:2011hd, Bhattacharyya:2008jc}. What is specific about $k=0 $ is the decoupling of time. 

The geometry of $\mathscr{I}^+$ is equipped with a conformal class of metrics rather than with a metric. 
From a representative of this class, we must be able to explore others by Weyl transformations, and this amounts to study conformal Carrollian geometry as opposed to plain Carrollian geometry (see \cite{Duval:2014uoa}). 

The action of Weyl transformations on the elements of the Carrollian geometry on a surface $\mathscr{S}$ is inherited from \eqref{conmet}:
 \begin{equation}
 \label{weyl-geometry}
a_{ij}\to\frac{a_{ij}}{{\cal B}^2}, \quad b_i\to \frac{b_i}{{\cal B}}, 
\quad 
\Omega \to \frac{\Omega}{{\cal B}},
\end{equation}
where $\mathcal{B}=\mathcal{B}(t,\mathbf{x})$ is an arbitrary function.  The Carrollian vorticity \eqref{carom} and shear \eqref{carshexp-tempcon} transform covariantly under  \eqref{weyl-geometry}: $
\varpi_{ij}\to \frac{1}{\mathcal{B}}\varpi_{ij}$,  $\xi_{ij}\to \frac{1}{\mathcal{B}}\xi_{ij}
$.
However, the Levi--Civita--Carroll covariant derivatives $\hat{\pmb{\nabla}}$ and $\hat\partial_t$ defined
previously for Carrollian geometry are not covariant under \eqref{weyl-geometry}. 
Following \cite{CMPPS1}, they must be replaced 
with
Weyl--Carroll covariant spatial and time derivatives built on the Carrollian acceleration $\varphi_i$ \eqref{caracc} and the Carrollian expansion \eqref{carshexp-tempcon},
which transform as connections:
 \begin{equation}
 \label{weyl-geometry-2-abs}
\varphi_{i}\to \varphi_{i}-\hat\partial_i\ln \mathcal{B},\quad \theta\to \mathcal{B}\theta-\frac{2}{\Omega}\partial_t \mathcal{B}.
\end{equation} 
In particular, these can be combined in\footnote{Contrary to $\varphi_{i}$, $\alpha_i$ is not a Carrollian one-form, \emph{i.e.} it does not transform covariantly under Carrollian diffeomorphisms \eqref{cardifs}.}
\begin{equation}
 \label{weyl-geometry-alpha}
\alpha_i=\varphi_{i}-\frac{\theta}{2}b_i,
\end{equation} 
transforming under Weyl rescaling as:
\begin{equation}
 \label{weyl-geometry-alpha-trans}
\alpha_i\to \alpha_{i}-\partial_i\ln \mathcal{B}.\end{equation} 

The Weyl--Carroll covariant derivatives $\hat{\mathscr{D}}_i$
and
$\hat{\mathscr{D}}_t$ are defined according to the pattern \eqref{Wconc}, \eqref{Wv}. They obey
\begin{equation}
\hat{\mathscr{D}}_j a_{kl}=0,\quad \hat{\mathscr{D}}_t a_{kl}=0.
\end{equation}
For a weight-$w$ scalar function  $\Phi$, or a weight-$w$ vector $V^i$,  \emph{i.e.}  scaling with $\mathcal{B}^w$ under \eqref{weyl-geometry}, 
we introduce
\begin{equation}
\label{CWs-Phivec}
\hat{\mathscr{D}}_j \Phi=\hat\partial_j \Phi +w \varphi_j \Phi,\quad 
\hat{\mathscr{D}}_j V^l=\hat\nabla_j V^l +(w-1) \varphi_j V^l +\varphi^l V_j -\delta^l_j V^i\varphi_i,
\end{equation}
which leave the weight unaltered.  Similarly,
we define 
\begin{equation}
\label{CWtimecovderscvec}
\frac{1}{\Omega}\hat{\mathscr{D}}_t \Phi=\frac{1}{\Omega}\hat\partial_t \Phi +\frac{w}{2} \theta\Phi=
\frac{1}{\Omega}\partial_t \Phi +\frac{w}{2} \theta \Phi,
\end{equation}
and 
\begin{equation}
\label{CWtimecovdervecform}
\frac{1}{\Omega}\hat{\mathscr{D}}_t V^l=\frac{1}{\Omega}\hat\partial_t V^l +\frac{w-1}{2} \theta V^l=
\frac{1}{\Omega}\partial_t V^l +\frac{w}{2} \theta V^l
+\xi^{l}_{\hphantom{l}i} V^i ,
\end{equation}
where  $\frac{1}{\Omega}\hat{\mathscr{D}}_t$ increases the weight by one unit.
The action of  $\hat{\mathscr{D}}_i$
and
$\hat{\mathscr{D}}_t$  on any other tensor is obtained using the Leibniz rule.

The Weyl--Carroll connection is torsion-free because
\begin{equation}
\label{CWcontor}
\left[\hat{\mathscr{D}}_i,\hat{\mathscr{D}}_j\right]\Phi=
\frac{2}{\Omega}\varpi_{ij}\hat{\mathscr{D}}_t\Phi
+w \left(\varphi_{ij}-\varpi_{ij} \theta\right)
\Phi
\end{equation}
does not contain terms of the type $\hat{\mathscr{D}}_k\Phi$. Here $\varphi_{ij}=\hat\partial_i\varphi_j-\hat\partial_j\varphi_i$ is a Carrollian two-form, not conformal though. Connection \eqref{CWcontor}
is accompanied with its own curvature tensors, which emerge in 
the commutation of Weyl--Carroll covariant derivatives acting \emph{e.g.} on vectors:
\begin{equation}
\label{CWcurvten}
\left[\hat{\mathscr{D}}_k,\hat{\mathscr{D}}_l\right]V^i=
\left( \hat{\mathscr{R}}^i_{\hphantom{i}jkl} - 2
\xi^{i}_{\hphantom{i}j}
\varpi_{kl} 
\right)
V^j+
\varpi_{kl}\frac{2}{\Omega}\hat{\mathscr{D}}_t V^i
+w \left(\varphi_{kl} -\varpi_{kl} \theta\right)
V^i.
\end{equation}
The combination $\varphi_{kl} -\varpi_{kl} \theta$ forms a weight-$0$ conformal two-form, whose dual $\ast\varphi-\ast\varpi\theta$ is conformal of weight $2$ ($\ast\varpi$ is defined in \eqref{starvarpi} and similarly $\ast\varphi=\frac{1}{2} \eta^{ij} \varphi_{ij}$). Moreover 
\begin{eqnarray}
\label{CWRiemann}
\hat{\mathscr{R}}^i_{\hphantom{i}jkl} &=&\hat r^i_{\hphantom{i}jkl}
-\delta^i_j\varphi_{kl}
-a_{jk} \hat{\nabla}_l \varphi^i
+a_{jl} \hat{\nabla}_k \varphi^i 
+\delta^i_k \hat{\nabla}_l \varphi_j 
-\delta^i_l \hat{\nabla}_k \varphi_j 
\nonumber
\\ 
&&+\varphi^i\left(\varphi_k a_{jl}-\varphi_l a_{jk}\right)
-\left(\delta^i_k a_{jl}-\delta^i_l a_{jk}\right)\varphi_m\varphi^m+
\left(\delta^i_k \varphi_l-\delta^i_l \varphi_k\right)\varphi_j
\end{eqnarray}
is the Riemann--Weyl--Carroll weight-$0$ tensor, from which we define
\begin{equation}
\label{CWricci}
\hat{\mathscr{R}}_{ij}=\hat{\mathscr{R}}^k_{\hphantom{k}ikj}=\hat r_{ij} 
+a_{ij} \hat{\nabla}_k \varphi^k
-\varphi_{ij}.
\end{equation}
We also quote
\begin{equation}
\left[\frac{1}{\Omega}\hat{\mathscr{D}}_{t},\hat{\mathscr{D}}_i\right]\Phi= w \hat{\mathscr{R}}_{i}\Phi-
\xi^{j}_{\hphantom{j}i}\hat{\mathscr{D}}_j^{\vphantom{j}} \Phi
\label{CWrsc}
\end{equation}
and 
\begin{equation}
\left[\frac{1}{\Omega}\hat{\mathscr{D}}_{t},\hat{\mathscr{D}}_i\right]V^i= 
(w-2)\hat{\mathscr{R}}_{i} V^i
-V^i\hat{\mathscr{D}}_j \xi^{j}_{\hphantom{j}i}
-\xi^{j}_{\hphantom{j}i}\hat{\mathscr{D}}_jV^i,
\label{CWrvec}
\end{equation}
with
\begin{equation}
\hat{\mathscr{R}}_{i}=\hat r_i
+\frac{1}{\Omega}\hat \partial_{t}\varphi_i
-\frac{1}{2} \hat\nabla_j\hat\gamma^{j}_{\hphantom{j}i}
+\xi^{j}_{\hphantom{j}i}\varphi_j=\frac{1}{\Omega} \partial_{t}\varphi_i-\frac{1}{2}\left(\hat \partial_i+\varphi_i\right)\theta.
\label{CWRvec}
\end{equation}
This is a Weyl-covariant weight-$1$ curvature one-form, where $\hat r_i$ is given in \eqref{carriemanntime}. 

The Ricci--Weyl--Carroll tensor \eqref{CWricci}
 is \emph{not} symmetric in general: $\hat{\mathscr{R}}_{ij}\neq \hat{\mathscr{R}}_{ji}$.  Using \eqref{carricci} we can recast it as
\begin{equation}
\label{CWricci-dec}
\hat{\mathscr{R}}_{ij}=\hat s_{ij}+\hat{\mathscr{K}} a_{ij}+\hat{\mathscr{A}} \eta_{ij},
\end{equation}
where we have introduced the Weyl-covariant scalar-electric and scalar-magnetic Gauss--Carroll curvatures
\begin{equation}
\label{CWscalar}
\hat{\mathscr{K}}=\frac{1}{2}a^{ij}\hat{\mathscr{R}}_{ij}=\hat{K}+ \hat{\nabla}_k \varphi^k
,\quad \hat{\mathscr{A}}=\frac{1}{2}\eta^{ij}\hat{\mathscr{R}}_{ij}=  \hat{A}- \ast \varphi
\end{equation}
both of weight $2$.

Before closing the present section, it is desirable to make a clarification: 
Weyl transformations \eqref{weyl-geometry} should not be confused with the action of the conformal Carroll group, which is a subset of Carrollian diffeomorphisms defined as\footnote{The subscript $2$ stands for level-$2$ conformal Carroll group. For a detailed discussion, see \cite{Duval:2014uva} .}
\begin{equation}
\textbf{CCarr}_2\left(\mathbb{R}\times \mathscr{S}, \text{d}\ell^2, \text{u}\right)=\left\{\phi\in \text{Diff}(\mathbb{R}\times \mathscr{S}),\quad \text{d}\ell^2\overset{\phi}{\longrightarrow}\text{e}^{-2 \Phi}\text{d}\ell^2\quad \text{u}\overset{\phi}{\longrightarrow}e^{\Phi}\text{u}\right\},
\label{CCarr}
\end{equation}  
where 
$\Phi\in\mathcal{C}^\infty(\mathbb{R}\times \mathscr{S})$, $\text{d}\ell^2$ is the spatial metric on $\mathscr{S}$ as in \eqref{dmet}, and $\text{u}=\frac{1}{\Omega}\partial_t$ the Carrollian time arrow.  This group is actually the zero-$k$ contraction of $\textbf{CIsom}\left(\mathscr{I}, \text{d}s^2 \right)$, the group of conformal isometries of the original finite-$k$ relativistic metric $\text{d}s^2$ on the boundary $\mathscr{I}$ of the corresponding AdS bulk:
\begin{equation}
\textbf{CIsom}\left(\mathscr{I}, \text{d}s^2\right)=\left\{\phi\in \text{Diff}(\mathscr{I}),\quad\text{d}s^2\overset{\phi}{\longrightarrow}\text{e}^{-2 \Phi}\text{d}s^2\right\}
\label{CIsom}
\end{equation}
with $ \Phi\in\mathcal{C}^\infty(\mathscr{I})$.
Indeed, consider the Lie algebra of conformal symmetries of $\text{d}s^2$, denoted $\mathfrak{cisom}\left(\mathscr{I}, \text{d}s^2\right)$ and spanned by vector fields $\text{X}=X^0\partial_0+X^i\partial_i$ such that 
\begin{equation}
\mathscr{L}_{\text{X}}\text{d}s^2=-2\lambda \text{d}s^2
\label{Liecisom}
\end{equation}
for some function $\lambda$ on $\mathscr{I}$. In order to perform the zero-$k$ contraction we write the generators as $\text{X}=k X^t\partial_0+X^i\partial_i$ (here $x^0=kt$, thus $X^0=kX^t$) and 
the metric $\text{d}s^2$ in the Randers--Papapetrou form \eqref{carrp}. 
At zero $k$ Eq. \eqref{Liecisom} splits into:\footnote{In coordinates, defining $\chi=\Omega X^t-b_jX^j$ , these equations are written as:
$$
\frac{1}{\Omega}\partial_t \chi + \varphi_jX^j=-\lambda, \quad \frac{1}{\Omega}\partial_t X^i
=0, \quad \hat\nabla^{(i}X^{j)}+\chi\left(
\xi^{ij}+\frac{1}{2}a^{ij}\theta\right)
=-\lambda a^{ij},
$$ 
which are manifestly covariant under Carrollian diffeomorphisms.}
\begin{equation}
\mathscr{L}_{\text{X}}\text{u}=\lambda\text{u},\quad\mathscr{L}_{\text{X}}\text{d}\ell^2=-2\lambda \text{d}\ell^2.
\label{Liecarr}
\end{equation} 
These are the equations the field $\text{X}$ must satisfy for belonging to $\mathfrak{ccarr}_2\left(\mathbb{R}\times \mathscr{S}, \text{d}\ell^2, \text{u}\right)$, the Lie algebra of the corresponding conformal Carroll group. This  confirms that
\begin{equation}
\textbf{CIsom}\left(\mathscr{I}, \text{d}s^2\right)\underset{k\to 0}{\longrightarrow}\textbf{CCarr}_2\left(\mathbb{R}\times \mathscr{S}, \text{d}\ell^2, \text{u}\right).
\label{CIsomtoCCarr}
\end{equation}
At last, if $\mathscr{S}$ is chosen to be the two-sphere and $\text{d}\ell^2$ the round metric, it can be shown (see \cite{Duval:2014uva}) that the corresponding conformal Carroll group is precisely the $\text{BMS}(4)$ group, which describes the asymptotic symmetries of an asymptotically flat $3+1$-dimensional  metric.

\subsection{Carrollian conformal fluid dynamics}\label{sec:carfluid}

\subsubsection*{Physical data and hydrodynamic equations}

More on the physics underlying the Carrollian limit can be found in 
\cite{CMPPS1}, with emphasis on hydrodynamics. This is precisely what we need here, since the original asymptotically AdS bulk Einstein spacetime is the holographic dual of a relativistic fluid hosted by its $2+1$-dimensional boundary. This relativistic fluid satisfying Eq. \eqref{T-cons}, will obey Carrollian dynamics at vanishing $k$. Even though the fluid has no velocity, it has non-trivial 
 hydrodynamics based on the following data:
\begin{itemize}
\item the energy density $\varepsilon(t,\mathbf{x})$ and the pressure $p(t,\mathbf{x})$, related here through a conformal equation of state   $
\varepsilon=2p$;
\item the heat currents $\pmb{Q}=Q_i(t,\mathbf{x})  \text{d}x^i$ and
$\pmb{\pi}=\pi_i(t,\mathbf{x})  \text{d}x^i$;
\item the viscous stress tensors 
 $\pmb{\Sigma}=\Sigma_{ij}(t,\mathbf{x})\,  \text{d}x^i\text{d}x^j$
 and
  $\pmb{\Xi}=\Xi_{ij}(t,\mathbf{x}) \text{d}x^i\text{d}x^j$.
\end{itemize}
The latter quantities are inherited from the relativistic ones (see \eqref{T}) as the following limits:
\begin{eqnarray}
\label{QexpC}
&Q_i=\lim\limits_{k\to 0}q_i,\quad
\pi_i=\lim\limits_{k\to 0}\frac{1}{k^2}\left(q_i-Q_i\right),
&
\\
\label{sigexpC}
&\Sigma_{ij}=-\lim\limits_{k\to 0}k^2\tau_{ij},\quad
\Xi_{ij}=-\lim\limits_{k\to 0}\left(\tau_{ij}+\frac{1}{k^2} \Sigma_{ij}\right).
&
\end{eqnarray}
Compared with the corresponding ones in the Galilean fluids, 
they are doubled because two orders seem to be required for describing the Carrollian dynamics. 
They obey 
\begin{equation}
\label{sigxi-symtr}
\Sigma_{ij}=\Sigma_{ji},\quad
\Sigma^{i}_{\hphantom{i}i}=0,\quad
\Xi_{ij}=\Xi_{ji}, \quad \Xi^i_{\hphantom{i}i}=0.
\end{equation}
The Carrollian energy and pressure are just the zero-$k$ limits of the corresponding relativistic quantities. In order to avoid symbols inflation, we have kept the same notation, $\varepsilon$ and $p$.

All these objects are Weyl-covariant with conformal weights $3$ for the pressure and energy density, $2$ for the heat currents, and $1$ for the viscous stress tensors (when all indices are lowered). 
They are well-defined in all examples we know from holography. Ultimately they should be justified  within a microscopic quantum/statistical approach, missing at present since 
the microscopic nature of a Carrollian fluid has not been investigated so far, except for 
\cite{CMPPS1}, where some elementary issues were addressed. 

Following this reference, the equations for a Carrollian fluid are as follows:
\begin{itemize}
\item a set of two scalar equations, both weight-$4$ Weyl-covariant:
\begin{eqnarray}
- \frac{1}{\Omega}\hat{\mathscr{D}}_t\varepsilon-\hat{\mathscr{D}}_i Q^{i}+\Xi^{ij}\xi_{ij}&=&0,
 \label{carE} 
\\
\label{carF} 
\Sigma^{ij}\xi_{ij}
&=&0;
\end{eqnarray}
\item two vector equations, Weyl-covariant of weight $3$:
\begin{eqnarray}
\hat{\mathscr{D}}_j p
+2Q^{i}\varpi_{ij}
+ \frac{1}{\Omega}\hat{\mathscr{D}}_t \pi_j
- \hat{\mathscr{D}}_i \Xi^i_{\hphantom{i}j}+ \pi_{i}\xi^{i}_{\hphantom{i}j}
&=& 0,
  \label{carG}\\
\frac{1}{\Omega}\hat{\mathscr{D}}_t 
Q_{j}- \hat{\mathscr{D}}_i \Sigma^{i}_{\hphantom{i}j}
+ Q_{i}\xi^{i}_{\hphantom{i}j}
&=&0.
 \label{carH} 
\end{eqnarray}
\end{itemize}
Equation \eqref{carE} is the energy conservation, whereas \eqref{carF} sets a geometrical constraint on the Carrollian viscous stress tensor $\Sigma_{ij}$. Equations \eqref{carG} and \eqref{carH} are dynamical equations involving the pressure $p=\nicefrac{\varepsilon}{2}$, the heat currents $Q_i$ and  $\pi_i$, and the viscous stress tensors $\Sigma_{ij}$ and $\Xi_{ij}$. { They are reminiscent of a momentum conservation, although somewhat degenerate due to the absence of fluid velocity.}

\subsubsection*{An example of Carrollian fluid}

The simplest non-trivial example of a Carrollian fluid is obtained as the Carrollian limit of the relativistic Robinson--Trautman fluid, studied at the end of Sec. \ref{resAdS} (see also \cite{Ciambelli:2017wou} and \cite{CMPPS1} for the relativistic and Carrollian approaches, respectively). 

The geometric Carrollian data are in this case 
\begin{equation}
\label{holoantimet}
\text{d}\ell^2=\frac{2}{P^2}\text{d}\zeta\text{d}\bar\zeta,
\end{equation}
$b_i=0$ and $\Omega = 1$. Hence the Carrollian shear vanishes ($\xi_{ij}=0$), whereas the expansion reads:
\begin{equation}
\label{carRT} 
\theta =-2 \partial_t \ln P.
\end{equation}
Similarly $\varpi_{ij}=0$, $\varphi_i=0$, $\varphi_{ij}=0$, and using results from App.  \ref{holo}, we find
\begin{equation}
\label{KART} 
\hat{\mathscr{K}}=2P^2 \partial_{\bar\zeta} \partial_\zeta\ln P,\quad 
\hat{\mathscr{A}}=0
\end{equation}
(in fact  $\hat{\mathscr{K}}= \hat{K}=K$), while
\begin{equation}
\label{RcurRT} 
\hat{\mathscr{R}}_{\bar\zeta}= \partial_{\bar\zeta}\partial_t\ln P,\quad 
\hat{\mathscr{R}}_{\bar\zeta}=  \partial_{\bar\zeta}\partial_t\ln P.
\end{equation}
From the relativistic heat current $\text{q}$ and viscous stress tensor $\tau$ displayed in \eqref{RT-q} and \eqref{RT-tau}, we obtain the Carrollian descendants:\footnote{Notice a useful identity:
$\partial_t\left(\frac{\partial^2_{\zeta} P}{P}\right) =
\frac{1}{P^2}\partial_{\zeta} \left(P^2\partial_t\partial_{\zeta}
\ln P
\right)$.\label{identity}}
\begin{eqnarray}
\label{QexpC-RT}
&\pmb{Q}= -\frac{1}{16\pi G}\left(\partial_{\zeta}K\text{d}\zeta+\partial_{\bar\zeta}K\text{d}\bar\zeta
\right),
\quad
\pmb{\pi}=0,
&
\\
\label{sigexpC-RT}
&\pmb{\Sigma}  = -\frac{1}{8\pi GP^2}
\left(
\partial_{\zeta} \left(P^2\partial_t\partial_{\zeta}
\ln P
\right)\text{d}\zeta^2+ 
\partial_{\bar\zeta} \left(P^2\partial_t\partial_{\bar\zeta}
\ln P
\right)
\text{d}\bar\zeta^2 \right),\quad
\pmb{\Xi}=0.
&
\end{eqnarray}
Due to the absence of shear, the hydrodynamic equation \eqref{carF} is identically satisfied, whereas 
\eqref{carE}, \eqref{carG}, \eqref{carH} are recast as:
\begin{eqnarray}
3\varepsilon\partial_t \ln P-\partial_t \varepsilon - \nabla_i Q^i&=&0,
 \label{carE2d} 
\\
 \partial_i p&=& 0,
  \label{carG2d}\\
\partial_t Q_i-2 Q_i \partial_t \ln P
-\nabla_j\Sigma^j_{\hphantom{j}i}&=&0.
 \label{carH2d} 
\end{eqnarray}
In agreement with the relativistic Robinson--Trautman fluid, the pressure $p$ (and so the energy density, since the fluid is conformal) must be space-independent. Furthermore, as expected from the relativistic case, Eq. \eqref{carH2d} is satisfied with $Q_i$ and $\Sigma_{ij}$ given in \eqref{QexpC-RT} and \eqref{sigexpC-RT}. Hence we are left with a single non-trivial equation, Eq. \eqref{carE2d}, the heat equation of the Carrollian fluid:
\begin{equation}
\label{RTcar}
3\varepsilon \partial_t \ln P-\partial_t \varepsilon+\frac{1}{16\pi G} \Delta K=0
\end{equation} 
with $\Delta= \nabla_j \nabla^j$ the Laplacian operator on $\mathscr{S}$.

Equation \eqref{RTcar} is exactly Robinson--Trautman's, Eq. \eqref{RT}. We note that the relativistic and the Carrolian dynamics lead to the same equations -- and hence to the same solutions $\varepsilon=\varepsilon(t)$. This is specific to the case under consideration, and it is actually expected since the bulk Einstein equations for a geometry with a shearless and vorticity-free null congruence lead to the Robinson--Trautman equation, irrespective of the presence of a cosmological constant, $\Lambda=-3k^2$: asymptotically locally AdS or locally flat spacetimes lead to the same dynamics. This is not the case in general though, because there is no reason for the relativistic dynamics to be identical to the Carrollian (see \cite{CMPPS1} for a detailed account of this statement). For example, when switching on more data, as in the case of the Pleba\'nski--Demia\'nski family, where all 
$b_i$, $\varphi_i$, $\varpi_{ij}$, as well as $\pi_i$ and $\Xi_{ij}$,
are on, the Carrollian equations are different from the relativistic ones. 

\section{The Ricci-flat limit \RNum{2}: derivative expansion and resummation}\label{sec:Ricci-recon}

We can summarize our observations as follows. Any four-dimensional Ricci-flat spacetime is associated with a two-dimensional spatial surface, emerging at null infinity and equipped with a conformal Carrollian geometry. This geometry is the host of a Carrollian fluid, obeying Carrollian hydrodynamics. Thanks to the relativistic-fluid/AdS-gravity duality, one can also safely claim that, conversely, any Carrollian fluid evolving on a spatial surface with Carrollian geometry is associated with a Ricci-flat geometry. This conclusion is reached by considering the simultaneous zero-$k$ limit of both sides of the quoted duality.
In order to make this statement operative, this limit must be performed inside the derivative expansion. When the latter is resummable in the sense discussed in Sec. \ref{resAdS}, the zero-$k$ limit will also affect the resummability conditions, and translate them in terms of Carrollian fluid dynamics.

\subsection{Back to the derivative expansion}

Our starting point is the derivative expansion of an asymptotically locally AdS spacetime, Eq. \eqref{papaefgenresc}.  The fundamental question is whether the latter admits a smooth zero-$k$ limit. 

We have implicitly assumed that the Randers--Papapetrou data of the three-dimensional pseudo-Riemannian conformal boundary $\mathscr{I}$ associated with the original Einstein spacetime, $a_{ij}$, $b_i$ and $\Omega$, remain unaltered at vanishing $k$, providing therefore directly the Carrollian data for the new spatial two-dimensional boundary $\mathscr{S}$ emerging at $\mathscr{I}^+$.\footnote{Indeed our ultimate goal is to set up a derivative expansion (in a closed resummed form under appropriate assumptions) for building up four-dimensional Ricci-flat spacetimes from a boundary Carrollian fluid, irrespective of its AdS origin. For this it is enough to assume $a_{ij}$, $b_i$ and $\Omega$ $k$-independent (as in \cite{CMPPS1}), and use these data as fundamental blocks for the Ricci-flat reconstruction. It should be kept in mind, however, that for general Einstein spacetimes, 
these may depend on $k$ with well-defined limit and subleading terms. Due to the absence of shear and to the particular structure of these solutions, the latter do not alter the Carrollian equations. This occurs for instance in Pleba\'nski--Demia\'nski  or in the Kerr--Taub--NUT sub-family, which will be discussed in Sec. \ref{perfluids}. In the following, we avoid discussing this kind of sub-leading terms, hence saving further technical developments.}  Following again the detailed analysis performed in \cite{CMPPS1}, we can match the various three-dimensional Riemannian quantities with the corresponding two-dimensional Carrollian ones:
\begin{equation}
\text{u}= -k^2\left(\Omega\text{d}t-\pmb{b}\right)
\end{equation}
and
\begin{equation}
\begin{array}{rcl}
\omega&=&\frac{k^2}{2}\varpi_{ij}\text{d}x^i\wedge\text{d}x^j,\\
\gamma&=&\ast \varpi,\\
\Theta&=&\theta,\\
\text{a}&=&k^2\varphi_i \text{d}x^i,\\
\text{A}&=&\alpha_i\text{d}x^i +\frac{\theta}{2}\Omega \text{d}t,\\
\sigma&=&\xi_{ij}\text{d}x^i\text{d}x^j,
\end{array}
\label{riemcar}
\end{equation}
where the left-hand-side quantities are Riemannian  
(given in Eqs. 
\eqref{omu},
\eqref{carexp},
\eqref{caracclim},
\eqref{A},
\eqref{sig}), and the right-hand-side ones Carrollian
(see
\eqref{caracc},
\eqref{carom},
\eqref{carshexp-tempcon},
\eqref{starvarpi}).

In the list \eqref{riemcar}, we have dealt with the first  derivatives, \emph{i.e.} connexion-related quantities.  We move now to second-derivative objects and collect the tensors relevant for the derivative expansion, following the same pattern (Riemannian vs. Carrollian):
\begin{eqnarray}
\label{Rlim}
\mathscr{R}&=&\frac{1}{k^2}\xi_{ij}\xi^{ij}+2\hat{\mathscr{K}}+2 k^2 \ast\varpi^2,
\\
\omega_\mu^{\hphantom{\mu}\lambda} \omega^{\vphantom{\lambda}}_{\lambda\nu}
\text{d}x^\mu\text{d}x^\nu&=& k^4
\varpi_i^{\hphantom{i}l} \varpi^{\vphantom{l}}_{lj}
\text{d}x^i\text{d}x^j,
\\
\omega^{\mu\nu} \omega_{\mu\nu}
&=& 2k^4\ast\varpi^2,
\\
\mathscr{D}_\nu\omega^{\nu}_{\hphantom{\nu}\mu} \text{d}x^\mu&=& k^2 
\hat{\mathscr{D}}_j\varpi^j_{\hphantom{j}i} \text{d}x^i
-2k^4\ast\varpi^2 \Omega\text{d}t+
2k^4\ast\varpi^2 \pmb{b}.
\end{eqnarray}
Using \eqref{S} this leads to 
\begin{equation}
\label{Slim}
\text{S}=-\frac{k^2}{2} \left(\Omega\text{d}t-\pmb{b}\right)^2 \xi_{ij}\xi^{ij} +k^4\pmb{s}-5 k^6 \left(\Omega\text{d}t-\pmb{b}\right)^2 \ast\varpi^2
\end{equation}
with the Weyl-invariant  tensor
\begin{equation}
\pmb{s}
=
2 \left(\Omega\text{d}t-\pmb{b}\right) \text{d}x^i  \eta^j_{\hphantom{j}i}
\hat{\mathscr{D}}_j \ast \varpi+\ast \varpi^2 \text{d}\ell^2-\hat{\mathscr{K}} \left(\Omega\text{d}t-\pmb{b}\right)^2.
\label{cars}
\end{equation}

In the derivative expansion \eqref{papaefgenresc}, two explicit  
divergences appear at vanishing $k$. The first 
originates from the first term of $\text{S}$, which is the shear contribution to the Weyl-covariant scalar curvature $\mathscr{R}$ of the three--dimensional  AdS boundary (Eq. \eqref{Rlim}).\footnote{This divergence is traced back in the Gauss--Codazzi equation relating the intrinsic and extrinsic curvatures of an embedded surface, to the intrinsic curvature of the host. When the size of a  fiber shrinks, the extrinsic-curvature contribution diverges.\label{sheardiv}} The second divergence comes from the Cotton tensor and is also due to the shear. It is fortunate -- and expected -- that counterterms coming from equal-order (non-explicitly written) $\sigma^2$ contributions, cancel out these singular terms.
This is suggestive that  \eqref{papaefgenresc} is well-behaved at zero-$k$, showing that the reconstruction of Ricci-flat spacetimes works starting from two-dimensional Carrollian fluid data. 

We will not embark here in proving finiteness at $k=0$, but rather confine our analysis to situations without shear, as we discussed already in Sec. \ref{resAdS} for Einstein spacetimes. Vanishing $\sigma$ in the pseudo-Riemannian boundary $\mathscr{I}$ implies indeed vanishing $\xi_{ij}$ in the Carrollian (see \eqref{riemcar}), and in this case, the divergent terms in $\text{S}$ and $\text{C}$ are absent.  Of course, other divergences may occur from higher-order terms in the derivative expansion. To avoid dealing with these issues, we will focus on the resummed version  of \eqref{papaefgenresc} \emph{i.e.} \eqref{papaefgenrescrec}, valid for algebraically special bulk geometries. This closed form is definitely smooth at zero $k$ and reads:
\begin{equation}
\boxed{
\text{d}s^2_{\text{res. flat}} =
-2\left(\Omega\text{d}t-\pmb{b}\right)\left(\text{d}r+r \pmb{\alpha}+\frac{r\theta \Omega}{2}\text{d}t\right)+r^2\text{d}\ell^2+\pmb{s}
+ \frac{\left(\Omega\text{d}t-\pmb{b}\right)^2}{\rho^2} \left(8\pi G \varepsilon r+
c \ast \varpi\right).}
\label{papaefresricf}
\end{equation}
Here
\begin{equation}
\label{rho2car}
 \rho^2= r^2 +\ast \varpi^2, 
\end{equation}  
$\text{d}\ell^2$,  $\Omega$, $\pmb{b}=b_i\text{d}x^i$, $\pmb{\alpha}=\alpha_i\text{d}x^i$, $\theta$ and $\ast \varpi$
are the Carrollian geometric objects introduced earlier, while $c$ and $\varepsilon$ are the zero-$k$ (finite) limits of the corresponding relativistic functions. Expression \eqref{papaefresricf} will grant by construction an exact Ricci-flat spacetime provided the conditions under which \eqref{papaefgenrescrec} was Einstein are fulfilled in the zero-$k$ limit. These conditions are the set of Carrollian hydrodynamic equations \eqref{carE},  \eqref{carF}, \eqref{carG} and \eqref{carH}, and  the integrability conditions, as they emerge from 
\eqref{heat-resum} and \eqref{visc-resum} at vanishing $k$. Making the latter explicit is the scope of next section.

Notice eventually that the Ricci-flat line element \eqref{papaefresricf} inherits Weyl invariance from its relativistic ancestor. The set of transformations \eqref{weyl-geometry}, \eqref{weyl-geometry-2-abs} and \eqref{weyl-geometry-alpha-trans}, supplemented with $\ast \varpi \to \mathcal{B}\ast\varpi$, $\varepsilon\to \mathcal{B}^3 \varepsilon$ and $c\to \mathcal{B}^3 c$, can indeed be absorbed by setting $r\to \mathcal{B}r$ ($\pmb{s}$ is Weyl invariant), resulting thus in the invariance of \eqref{papaefresricf}. In the relativistic case this invariance was due to the AdS conformal boundary. In the case at hand, this is rooted to the location of the two-dimensional spatial boundary $\mathscr{S}$ at null infinity $\mathscr{I}^+$.

\subsection{Resummation of the Ricci-flat derivative expansion}
\label{sec:resum-riccif}

\subsubsection*{The Cotton tensor in Carrollian geometry}

The Cotton tensor monitors from the boundary the global asymptotic structure of the  bulk four-dimensional Einstein spacetime (for higher dimensions, the boundary Weyl tensor is also involved, see footnote \ref{marie}).
 In order to proceed with our resummability analysis, we need to describe the zero-$k$ limit of the Cotton tensor \eqref{cotdef} and of its conservation equation \eqref{C-cons}. 

As already mentioned, at vanishing $k$ divergences do generally appear for some components of the Cotton tensor. These divergences are no longer present when \eqref{shear3d} is satisfied (see footnote \ref{sheardiv}), \emph{i.e.} in the absence of shear, which is precisely the assumption under which we are working with \eqref{papaefresricf}. Every piece of the three-dimensional relativistic Cotton tensor appearing in \eqref{C} has thus a well-defined limit. We therefore introduce
\begin{eqnarray}
\label{CexpC}
&\chi_i=\lim\limits_{k\to 0}c_i,\quad
\psi_i=\lim\limits_{k\to 0}\frac{1}{k^2}\left(c_i-\chi_i\right),
&
\\
\label{ChiexpC}
&X_{ij}=\lim\limits_{k\to 0}c_{ij},\quad
\Psi_{ij}=\lim\limits_{k\to 0}\frac{1}{k^2}\left(c_{ij}- X_{ij}\right).
&
\end{eqnarray}
The time components $c_0$, $c_{00}$ and $c_{0i}=c_{i0}$ vanish already at finite $k$ (due to \eqref{cotrans}), and  $\chi_i$,  $\psi_i$, $X_{ij}$ and $\Psi_{ij}$ are thus genuine Carrollian tensors transforming covariantly under Carrollian diffeomorphisms. Actually, in the absence of shear the Cotton current and stress tensor are given exactly (\emph{i.e.} for finite $k$) by $c_i=\chi_i+k^2  \psi_i$ and 
$c_{ij}=X_{ij}+k^2 \Psi_{ij}$. 

The scalar $c(t,\mathbf{x})$ is  Weyl-covariant of weight $3$ (like the energy density). As expected, it is expressed in terms of geometric Carrollian objects built on third-derivatives of the two-dimensional metric $\text{d}\ell^2$, $b_i$ and $\Omega$:
\begin{equation}
\label{c-Carrol}
c=\left(\hat{\mathscr{D}}_l\hat{\mathscr{D}}^l+2\hat{\mathscr{K}}
\right)\ast \varpi.
\end{equation}  
Similarly, the forms $\chi_i$ and  $\psi_i$, of weight $2$, are
\begin{eqnarray}
\label{chi-f-Carrol}
\chi_j&=&\frac{1}{2}\eta^l_{\hphantom{l}j}\hat{\mathscr{D}}_l\hat{\mathscr{K}}+ \frac{1}{2} \hat{\mathscr{D}}_j\hat{\mathscr{A}}-2\ast \varpi\hat{\mathscr{R}}_j,
\\
\label{psi-f-Carrol}
\psi_j&=&3\eta^l_{\hphantom{l}j}\hat{\mathscr{D}}_l\ast \varpi^2.
\end{eqnarray}  
Finally, the weight-$1$ symmetric and traceless rank-two tensors read:
\begin{eqnarray}
\label{X-2-Carrol}
X_{ij}&=&\frac{1}{2}\eta^l_{\hphantom{l}j}\hat{\mathscr{D}}_l
\hat{\mathscr{R}}_i+
\frac{1}{2} \eta^l_{\hphantom{l}i}\hat{\mathscr{D}}_j
\hat{\mathscr{R}}_l,
\\
\label{Psi-2-Carrol}
\Psi_{ij}&=&\hat{\mathscr{D}}_i \hat{\mathscr{D}}_j\ast \varpi -\frac{1}{2}a_{ij} \hat{\mathscr{D}}_l \hat{\mathscr{D}}^l \ast \varpi -\eta_{ij} \frac{1}{\Omega}  \hat{\mathscr{D}}_t\ast \varpi^2.
\end{eqnarray}  
Observe that $c$ and the subleading terms $\psi_i$ and $\Psi_{ij}$ are present only when the vorticity is non-vanishing ($\ast\varpi\neq 0$). All these are of gravito-magnetic nature.

The tensors $c$, $\chi_{i}$, $\psi_i$, $X_{ij}$ and $\Psi_{ij}$ should be considered as the two-dimensional Carrollian resurgence of the three-dimensional Riemannian Cotton tensor. They should be referred to as Cotton descendants  (there is no Cotton tensor in two dimensions anyway), and obey identities inherited at zero $k$ from its conservation equation.\footnote{Observe that the Cotton tensor enters in Eq. \eqref{eqn:Tref} with an extra factor $\nicefrac{1}{k}$, the origin of which is explained in  footnote \ref{k-fac}. Hence, the advisable prescription is to analyze the small-$k$ limit of   $\frac{1}{k}\nabla^\mu C_{\mu\nu}=0$.} These are similar to the hydrodynamic equations \eqref{carE},  \eqref{carF}, \eqref{carG} and \eqref{carH}, satisfied by the different pieces of the energy--momentum tensor $\varepsilon$, $Q_i$, $\pi_i$, $\Sigma_{ij}$ and $\Xi_{ij}$, and translating its conservation. In the case at hand, the absence of shear trivializes \eqref{carF} and discards the last term in the other three equations:   
\begin{eqnarray}
\frac{1}{\Omega}\hat{\mathscr{D}}_t c+\hat{\mathscr{D}}_i \chi^{i}
&=&0,
 \label{carEcot} 
\\
\frac{1}{2}\hat{\mathscr{D}}_j c
+2\chi^{i}\varpi_{ij}
+ \frac{1}{\Omega}\hat{\mathscr{D}}_t \psi_j
- \hat{\mathscr{D}}_i \Psi^i_{\hphantom{i}j}
&=& 0,
  \label{carGcot}\\
\frac{1}{\Omega}\hat{\mathscr{D}}_t 
\chi_{j}- \hat{\mathscr{D}}_i X^{i}_{\hphantom{i}j}
&=&0.
 \label{carHcot} 
\end{eqnarray}
One appreciates from these equations why it is important to keep the subleading corrections at vanishing $k$, both in the Cotton current $c_\mu$ and in the Cotton stress tensor  $c_{\mu\nu}$. As for the energy--momentum tensor, ignoring them would simply lead to wrong Carrollian dynamics.

\subsubsection*{The resummability conditions}

We are now ready to address the problem of resummability in Carrollian framework, for Ricci-flat spacetimes. In the relativistic case, where one describes relativistic hydrodynamics on the pseudo-Riemannian boundary of an asymptotically locally AdS spacetime, 
resummability -- or integrability -- equations are Eqs. \eqref{heat-resum} and \eqref{visc-resum}. These determine the friction components of the fluid energy--momentum tensor in terms of geometric data, captured by the Cotton tensor (current and stress components),
via a sort of gravitational electric--magnetic duality, transverse to the fluid congruence. Equipped with those, the fluid equations \eqref{T-cons} guarantee that the bulk is Einstein, \emph{i.e.} that bulk Einstein equations are satisfied.

Correspondingly, using \eqref{QexpC}, \eqref{sigexpC}, \eqref{CexpC} and \eqref{ChiexpC},
the zero-$k$ limit of Eq. \eqref{heat-resum} sets up a duality relationship among the Carrollian-fluid heat current $Q_i$ and the Carrollian-geometry third-derivative vector $\chi_i$:
\begin{equation}
\boxed{
Q_i=\frac{1}{8\pi G}\eta^j_{\hphantom{j}i}\chi_j\\
=-
\frac{1}{16\pi G}\left(\hat{\mathscr{D}}_i\hat{\mathscr{K}}- \eta^j_{\hphantom{j}i} \hat{\mathscr{D}}_j\hat{\mathscr{A}}+4\ast \varpi  \eta^j_{\hphantom{j}i} \hat{\mathscr{R}}_j\right),
}
\label{heat-resum-car-Q}
\end{equation}
while Eqs. \eqref{visc-resum} allow to relate the Carrollian-fluid quantities $\Sigma_{ij}$ and $\Xi_{ij}$, to the Carrollian-geometry ones $X_{ij}$ and $\Psi_{ij}$:
\begin{equation}
\boxed{
\Sigma_{ij}=\frac{1}{8\pi G}\eta^l_{\hphantom{l}i}X_{lj}=\frac{1}{16\pi G}\left(
\eta^k_{\hphantom{k}j}
\eta^l_{\hphantom{l}i}
\hat{\mathscr{D}}_k
\hat{\mathscr{R}}_l-
\hat{\mathscr{D}}_j
\hat{\mathscr{R}}_i
\right),
}
\label{visc-resum-car-sig}
\end{equation}
and
\begin{equation}
\boxed{
\Xi_{ij}=\frac{1}{8\pi G}\eta^l_{\hphantom{l}i}\Psi_{lj}=\frac{1}{8\pi G}\left(
\eta^l_{\hphantom{l}i}
\hat{\mathscr{D}}_l \hat{\mathscr{D}}_j\ast \varpi +\frac{1}{2}\eta_{ij} \hat{\mathscr{D}}_l \hat{\mathscr{D}}^l \ast \varpi -a_{ij} \frac{1}{\Omega}  \hat{\mathscr{D}}_t\ast \varpi^2
\right).
}
\label{visc-resum-car-xi}
\end{equation}
One readily shows that \eqref{sigxi-symtr} is satisfied as a consequence of the symmetry and tracelessness of $X_{ij}$ and $\Psi_{ij}$.

One can finally recast the Carrollian hydrodynamic equations  \eqref{carE},  \eqref{carF}, \eqref{carG} and \eqref{carH} for the fluid under consideration. Recalling that the shear is assumed to vanish,
\begin{equation}
\label{xi}
\xi_{ij}=\frac{1}{2\Omega}\left(\partial_t a_{ij}-a_{ij}\partial_t \ln \sqrt{a}\right)=0,
\end{equation}
Eq. \eqref{carF} is trivialized. Furthermore, Eq.  \eqref{carH} is automatically satisfied with $Q_j$ and $\Sigma^i_{\hphantom{i}j}$ given above, thanks also to Eq. \eqref{carHcot}. We are therefore left with two equations for the energy density $\varepsilon$ and the heat current $\pi_i$: 
\begin{itemize}
\item one scalar equation from \eqref{carE}:
\begin{equation}
\boxed{
- \frac{1}{\Omega}\hat{\mathscr{D}}_t\varepsilon+\frac{1}{16\pi G}\hat{\mathscr{D}}^i\left(\hat{\mathscr{D}}_i\hat{\mathscr{K}}- \eta^j_{\hphantom{j}i} \hat{\mathscr{D}}_j\hat{\mathscr{A}}+4\ast \varpi  \eta^j_{\hphantom{j}i} \hat{\mathscr{R}}_j\right)=0;
}
\label{heat-eq-carE}
\end{equation}
\item one vector equation from \eqref{carG}: 
\begin{equation}
\boxed{
\hat{\mathscr{D}}_j \varepsilon
+4 \ast \varpi \eta^i_{\hphantom{i}j} Q_{i}
+ \frac{2}{\Omega}\hat{\mathscr{D}}_t \pi_j
-2 \hat{\mathscr{D}}_i \Xi^i_{\hphantom{i}j}
= 0}
  \label{mom-eq-carG}
  \end{equation}
with $ Q_{i}$ and $\Xi^i_{\hphantom{i}j}$ given in \eqref{heat-resum-car-Q} and \eqref{visc-resum-car-xi}.  
\end{itemize}

These last two equations are Carrollian equations, describing time and space evolution of the fluid energy and heat current, as a consequence of transport phenomena like heat conduction and friction.  These phenomena have been identified by duality to geometric quantities, and one recognizes distinct gravito-electric  (like 
$\hat{\mathscr{K}}$) and gravito-magnetic contributions  (like $\hat{\mathscr{A}}$). It should also be stressed that not all the terms are independent and one can reshuffle them using identities relating the Carrollian curvature elements. In the absence of shear, \eqref{cars-van} holds and all information about
$\hat{\mathscr{R}}_{ij}$ in \eqref{CWricci-dec} is stored in $\hat{\mathscr{K}}$ and $\hat{\mathscr{A}}$, while other geometrical data are supplied by $\hat{\mathscr{R}}_{i}$ in \eqref{CWRvec}. All these obey
\begin{equation}
\begin{array}{rcl}
 \frac{2}{\Omega}\hat{\mathscr{D}}_t \ast \varpi +\hat{\mathscr{A}}&=&0
,\\
 \frac{1}{\Omega}\hat{\mathscr{D}}_t \hat{\mathscr{K}}
-a^{ij}\hat{\mathscr{D}}_i  \hat{\mathscr{R}}_{j}  &=& 0 ,
\\
\frac{1}{\Omega}\hat{\mathscr{D}}_t \hat{\mathscr{A}}+ \eta^{ij}\hat{\mathscr{D}}_i  \hat{\mathscr{R}}_{j} &=&0
,
\end{array}
\label{Carroll-Bianchi}
\end{equation}
which originate from three-dimensional Riemannian Bianchi identities and emerge along the $k$-to-zero limit.  

\subsubsection*{Summarizing}

Our analysis of the zero-$k$ limit in the derivative expansion \eqref{papaefgenrescrec}, valid assuming the absence of shear, has the following salient features.
\begin{itemize}
\item As the general derivative expansion \eqref{papaefgenresc}, this limit reveals a two-dimensional spatial boundary $\mathscr{S}$ located at   $\mathscr{I}^+$. It is endowed with a Carrollian geometry, encoded in $a_{ij}$, $b_i$ and $\Omega$, all functions of $t$ and $\mathbf{x}$. This is inherited from the conformal three-dimensional pseudo-Riemannian boundary $\mathscr{I}$ of the original Einstein space. 
\item The Carrollian boundary $\mathscr{S}$ is the host of a Carrollian fluid, obtained as the limit of a relativistic fluid, and described in terms of its energy density $\varepsilon$, and its friction tensors  $Q_i$, $\pi_i$, $\Sigma_{ij}$ and $\Xi_{ij}$. \item When the friction tensors  $Q_i$, $\Sigma_{ij}$ and $\Xi_{ij}$ of the Carrollian fluid are given in terms of the geometric objects 
$\chi_i$, $X_{ij}$ and $\Psi_{ij}$ using \eqref{heat-resum-car-Q}, \eqref{visc-resum-car-sig} and \eqref{visc-resum-car-xi}, and when the energy density $\varepsilon$ and the current $\pi_i$ obey the hydrodynamic equations  \eqref{heat-eq-carE} and   \eqref{mom-eq-carG}, the limiting resummed derivative expansion \eqref{papaefresricf} is an exact Ricci-flat spacetime.
\item The bulk spacetime is in general asymptotically locally flat. 
This property is encoded in the zero-$k$ limit of the Cotton tensor, \emph{i.e.} in the 
Cotton Carrollian descendants $c$, $\chi_i$ and $X_{ij}$. 

\end{itemize}

The bulk Ricci-flat spacetime obtained following the above procedure is algebraically special. We indeed observe that the bulk congruence $\partial_r$ is null. Moreover, it is geodesic and shear-free. 
To prove this last statement, we rewrite the metric \eqref{papaefresricf} in terms of a null tetrad $(\mathbf{k},\mathbf{l},\mathbf{m},\bar{\mathbf{m}})$:
\begin{equation}
\text{d}s^2_{\text{res. flat}} =-2\mathbf{k}\mathbf{l}+2\mathbf{m}\bar{\mathbf{m}}\,,\quad
\mathbf{k}\cdot\mathbf{l}=-1\,,\quad\mathbf{m}\cdot\bar{\mathbf{m}}=1\,,
\end{equation}
where $\mathbf{k}=-\left(\Omega\text{d}t-\pmb{b}\right)$ is the dual of $\partial_r$ and
\begin{equation}
\mathbf{l}=-\text{d}r-r \pmb{\alpha}-\frac{r\theta \Omega}{2}\text{d}t+
\frac{\pmb{\psi}}{6\ast\varpi}
+\frac{\Omega\text{d}t-\pmb{b}}{2\rho^2} \left(8\pi G \varepsilon r+c \ast \varpi-\rho^2\hat{\mathscr{K}}\right)\,,
\end{equation}
(here $\pmb{\psi}=\psi_i \text{d}x^i$), along with
\begin{equation}
2\mathbf{m}\bar{\mathbf{m}}=\rho^2\text{d}\ell^2\,.
\end{equation}
Using the above results and repeating the analysis of App. A.2 in \cite{Petropoulos:2015fba}, we find that $\partial_r$ is shear-free
due to \eqref{xi}.

According to the Goldberg--Sachs theorem, the bulk spacetime \eqref{papaefresricf} is therefore of Petrov type II, III, D, N or O. The precise type is encoded in the Carrollian tensors $\varepsilon^\pm$, $Q^\pm_i$ and $\Sigma^\pm_{ij}$
\begin{equation}
\begin{array}{rcl}
\varepsilon^\pm &=&\varepsilon\pm\frac{\text{i}}{8\pi G} c, \\
Q^\pm_{i} &=&Q_{i}\pm\frac{\text{i}}{8\pi G} \chi_{i} ,\\
\Sigma^\pm_{ij} &=&\Sigma_{ij}\pm\frac{\text{i}}{8\pi G} X_{ij}.
\end{array}
\label{principal}
\end{equation}
Working again in holomorphic coordinates, we find the compact result
 \begin{eqnarray}
 \pmb{Q}^+
 &=& \frac{\text{i}}{4\pi G}\chi_\zeta \text{d}\zeta,
\label{principal-holoQ}
 \\
 \pmb{\Sigma}^+
&=& \frac{\text{i}}{4\pi G}X_{\zeta\zeta} \text{d}\zeta^2,
\label{principal-holosig}
\end{eqnarray}
and their complex-conjugates $\pmb{Q}^-$ and $ \pmb{\Sigma}^-$. 
These Carrollian geometric tensors encompass the information on the canonical complex functions describing the Weyl-tensor decomposition in terms of principal null directions -- usually referred to as $\Psi_a, a=0,\ldots,4$.

\section{Examples}\label{sec:ex}

There is a plethora of Carrollian fluids that can be studied. We will analyze here the class of \emph{perfect conformal fluids}, and will complete the discussion of Sec. \ref{sec:carfluid} on the \emph{Carrollian Robinson--Trautman fluid}. In each case, assuming the integrability conditions \eqref{heat-resum-car-Q}, \eqref{visc-resum-car-sig} and \eqref{visc-resum-car-xi} are fulfilled and the hydrodynamic equations \eqref{heat-eq-carE} and   \eqref{mom-eq-carG}
are obeyed, a Ricci-flat spacetime is reconstructed from the  Carrollian spatial boundary $\mathscr{S}$
at  $\mathscr{I}^+$. More examples exist like the Pleba\'nski--Demia\'nski or the Weyl axisymmetric solutions, assuming extra symmetries (but not necessarily stationarity) for a viscous Carrollian fluid. These would require a more involved presentation.

\subsection{Stationary Carrollian perfect fluids and Ricci-flat Kerr--Taub--NUT families}
\label{perfluids}

We would like to illustrate our findings and reconstruct from purely Carrollian fluid dynamics the family of Kerr--Taub--NUT stationary Ricci-flat black holes.  We pick for that the following geometric data: $a_{ij}(\mathbf{x})$, $b_{i}(\mathbf{x})$ and $\Omega = 1$. Stationarity is implemented in these fluids by requiring that all the quantities involved are time independent.

Under this assumption, the Carrollian shear $\xi_{ij}$ vanishes together with the Carrollian expansion $\theta$, whereas constant $\Omega$ makes the Carrollian acceleration $\varphi_i$ vanish as well (Eq. \eqref{caracc}). Consequently 
\begin{equation}
\hat{\mathscr{A}}=0,\quad 
\hat{\mathscr{R}}_i =0, 
\end{equation}
and 
we are left with non-trivial curvature and vorticity:
\begin{equation}
\label{K-carvort-perf}
\hat{\mathscr{K}}=\hat{K}=K,
\quad\varpi_{ij}=\partial_{[i}b_{j]}=\eta_{ij}\ast \varpi .
\end{equation}
The Weyl--Carroll spatial covariant derivative $\hat{\mathscr{D}}_i$ reduces to the ordinary covariant derivative $\nabla_i$, whereas the action of the Weyl--Carroll temporal covariant derivative $\hat{\mathscr{D}}_t$ vanishes.

We further assume that the Carrollian fluid is perfect:  $Q_i$, $\pi_i$, $\Sigma_{ij}$ and $\Xi_{ij}$ vanish. This assumption is made according to the pattern of Ref. \cite{Mukhopadhyay:2013gja}, where the asymptotically AdS Kerr--Taub--NUT spacetimes were studied starting from relativistic perfect fluids. Due to the duality relationships \eqref{heat-resum-car-Q}, \eqref{visc-resum-car-sig} and \eqref{visc-resum-car-xi}  among the friction tensors of the Carrollian fluid and the geometric quantities $\chi_i$, $X_{ij}$ and $\Psi_{ij}$, the latter must also vanish. Using \eqref{chi-f-Carrol}, \eqref{X-2-Carrol} and \eqref{Psi-2-Carrol},
this sets the following simple geometric constraints:
\begin{equation}
\label{perf-con}
\chi_i=0\Leftrightarrow \partial_i K =0,
\end{equation}
and
\begin{equation}
\label{perf-varpi}
\Psi_{ij}=0\Leftrightarrow \left(\nabla_i \nabla_j-\frac{1}{2}a_{ij} \nabla_l \nabla^l\right) \ast \varpi =0,
\end{equation}
whereas $X_{ij}$ vanishes identically without bringing any further restriction.
These are equations for the metric $a_{ij}(\mathbf{x})$ and the scalar vorticity $ \ast \varpi $, from which we can extract $b_{i}(\mathbf{x})$.  
Using \eqref{c-Carrol}, we also learn that 
\begin{equation}
\label{perf-c}
c=\left(
 \Delta
+2K\right) \ast \varpi,
\end{equation}
where  $\Delta= \nabla_l \nabla^l$ is the ordinary Laplacian operator on $\mathscr{S}$. 
The last piece of the geometrical data, \eqref{psi-f-Carrol}, it is non-vanishing and reads:
\begin{equation}
\psi_j=3\eta^l_{\hphantom{l}j}\partial_l\ast \varpi^2.
\end{equation}
Finally, we must impose the fluid equations  \eqref{heat-eq-carE} and   \eqref{mom-eq-carG}, leading to 
\begin{equation}
\label{hydro-perf}
\partial_t \varepsilon=0,\quad \partial_i \varepsilon=0. 
\end{equation}
The energy density $ \varepsilon$ of the Carrollian fluid is therefore a constant, which will be identified to the bulk mass parameter  $M = 4\pi G \varepsilon$.

Every stationary Carrollian geometry encoded in $a_{ij}(\mathbf{x})$ and $b_{i}(\mathbf{x})$
with constant scalar curvature  $ K $  hosts a conformal Carrollian perfect fluid with constant energy density, and is associated with the following exact Ricci-flat spacetime: 
\begin{equation}
\text{d}s^2_{\text{perf. fl.}}=
-2
\left(\text{d}t-\pmb{b}\right)\text{d}r
+\frac{2M r+c\ast\varpi-K\rho^2}{\rho^2}  \left(\text{d}t-\pmb{b}\right)^2
+\left(\text{d}t-\pmb{b}\right)\frac{\pmb{\psi}}{3\ast\varpi}
+\rho^2\text{d}\ell^2,
\label{papaefresricfKTN}
\end{equation}
where $\rho^2=r^2+\ast\varpi^2$. The vorticity $\ast\varpi$ is determined by Eq. \eqref{perf-varpi}, solved on a constant-curvature background. 

Using holomorphic coordinates (see App. \ref{holo}), a constant-curvature metric on $\mathscr{S}$ reads:
\begin{equation}
\label{CF-perfect}
\text{d}\ell^2=\frac{2}{P^2}\text{d}\zeta\text{d}\bar\zeta
\end{equation}
with 
\begin{equation}
\label{constK}
P= 1+\frac{K}{2}\zeta \bar \zeta,\quad K=0,\pm 1,
\end{equation}
corresponding to $S^2$ and $E_2$ or $H_2$ (sphere and Euclidean or hyperbolic planes). 
Using these expressions we can integrate \eqref{perf-varpi}. The general solution depends on three real, arbitrary parameters, $n$, $a$ and $\ell$:
\begin{equation}
\label{star-varpi-perf}
\ast\varpi=n+a-\frac{2a}{P} + \frac{\ell}{P} \left(1-\vert K\vert\right)\zeta\bar\zeta .
\end{equation}
The parameter $\ell$ is relevant in the flat case exclusively. We can further integrate \eqref{carom} and find thus
\begin{equation}
\label{b-perf}
\pmb{b} = \frac{\text{i}}{P}\left(
n-\frac{a}{P} +\frac{\ell}{2P} \left(1-\vert K\vert\right)\zeta\bar\zeta 
\right)
\left(
\bar\zeta\text{d}\zeta
-\zeta\text{d}\bar\zeta 
\right).
\end{equation}
It is straightforward to determine the last pieces entering the bulk resumed metric \eqref{papaefresricfKTN}:
\begin{equation}
\label{c-perf}
c=2Kn +2\ell  \left(1-\vert K\vert\right)
 \end{equation}
and 
\begin{equation}
\label{omeu-perf}
\frac{\pmb{\psi}}{3\ast\varpi}
=2\eta^j_{\hphantom{l}i} \partial_j\ast \varpi \text{d}x^i =
2\text{i}\frac{Ka+\ell 
\left(1-\vert K\vert\right)}{P^2}
\left(
\bar\zeta\text{d}\zeta
-\zeta\text{d}\bar\zeta 
\right).
\end{equation}

In order to reach a more familiar form for the line element \eqref{papaefresricfKTN}, it is convenient to trade the complex-conjugate coordinates $\zeta$ and $\bar{\zeta}$ for their modulus\footnote{
The modulus and its range depend on the curvature. It is commonly expressed as:
$Z=\sqrt{2}\tan \frac{\varTheta}{2}$, ${0<\varTheta<\pi}$ for $S^2$;
$Z=\frac{R}{\sqrt{2}}$, $0<R<+\infty$ for $E_2$;  $Z=\sqrt{2}\tanh \frac{\Psi}{2}$,  
$0<\Psi<+\infty$ for $H_2$.\label{zetazetabarfam}} and argument
\begin{equation}
\zeta=Z \text{e}^{i\varPhi},\label{zeta}
\end{equation}
and move from Eddington--Finkelstein to  Boyer--Lindquist by setting 
\begin{equation}
\text{d}t\to\text{d}t-\frac{r^2+(n-a)^2}{\Delta_r}\text{d}r\,,\quad \text{d}\varPhi\to\text{d}\varPhi-\frac{Ka+\ell(1-\vert K\vert)}{\Delta_r}\text{d}r
\end{equation}
with 
\begin{equation}
\Delta_r=-2M r+K\left(r^2+a^2-n^2\right)+2\ell(n-a)(|K|-1).\label{Delta}
\end{equation}
We obtain finally:
\begin{eqnarray}
\nonumber
\text{d}s^2_{\text{perf. fl.}}&=&-\frac{\Delta_r}{\rho^2}\left(\text{d}t+ \frac{2}{P}\left(
n-\frac{a}{P} +\frac{\ell}{2P} \left(1-\vert K\vert\right)Z^2 
\right)Z^2
\text{d}\varPhi\right)^2+\frac{\rho^2}{\Delta_r}\text{d}r^2\\
&&+
\frac{2\rho^2}{P^2}\text{d}Z^2
+\frac{2Z^2}{\rho^2P^2}
\left(\left(K a+\ell\left(1-\vert K\vert\right)\right)\text{d}t-\left(r^2+\left(n-a\right)^2\right)\text{d}\varPhi\right)^2
\label{KTNBL}
\end{eqnarray}
with
\begin{equation}
P= 1+\frac{K}{2}Z^2,
\quad
\rho^2=r^2+\left(n+a-\frac{2a}{P} + \frac{\ell}{P} \left(1-\vert K\vert\right)Z^2\right)^2.
\end{equation}
This bulk metric is Ricci-flat for any value of the parameters $M$, $n$, $a$ and $\ell$ with $K=0,\pm 1$. For vanishing $n$, $a$ and $\ell$, and with $M>0$ and $K=1$, one recovers the standard asymptotically flat Schwarzschild solution with spherical horizon. For $K=0$ or $-1$, this is no longer Schwarzschild, but rather a metric belonging to the A class (see e.g. \cite{Podolsky}). 
The parameter $a$ switches on rotation, while $n$ is the standard nut charge. The parameter $\ell$ is also a rotational parameter available only in the flat-$\mathscr{S}$ case. Scanning over all these parameters, in combination with the mass and $K$, we recover the whole Kerr--Taub--NUT family of black holes, plus other, less familiar configurations, like  the A-metric quoted above. 

For the solutions at hand, the only potentially non-vanishing Carrollian boundary Cotton descendants are $c$ and $\pmb{\psi}$, displayed in \eqref{c-perf} and  \eqref{omeu-perf}. The first is non-vanishing for asymptotically locally flat spacetimes, and this requires non-zero $n$ or $\ell$. The second measures the bulk twist. In every case the metric \eqref{KTNBL} is Petrov type D.

We would like to conclude the example of Carrollian conformal perfect fluids with a comment regarding the isometries of the associated resummed Ricci-flat spacetimes with line element \eqref{KTNBL}. For vanishing  $a$ and $\ell$, there are  
\emph{four} isometry generators and the field is in this case a stationary gravito-electric and/or gravito-magnetic monopole (mass  and nut  parameters $M$, $n$). Constant-$r$ hypersurfaces are homogeneous spaces in this case.
The number  of Killing fields
is reduced to \emph{two} ($\partial_t$ and $\partial_\varPhi$) whenever any of the rotational parameters   $a$ or $\ell$ is non-zero. These parameters make the gravitational field dipolar. 

The bulk isometries are generally inherited  from the boundary symmetries, 
\emph{i.e.} the symmetries of the Carrollian geometry and the Carrollian fluid. The time-like Killing field $\partial_t$ is clearly rooted to the stationarity of the boundary data. The space-like ones have legs on $\partial_\varPhi$ and $\partial_Z$, and are associated to further boundary symmetries. 
From a Riemannian viewpoint, the metric \eqref{CF-perfect} with \eqref{constK} on the two-dimensional boundary surface $\mathscr{S}$ 
admits three Killing vector fields: 
\begin{eqnarray}
\label{XK}
\pmb{X}_1&=&\text{i}\left(\zeta \partial_\zeta-
\bar\zeta \partial_{\bar\zeta}
\right),\\
\label{YK}
\pmb{X}_2&=&\text{i}\left(\left(1-\frac{K}{2}\zeta^2\right)
\partial_\zeta-
\left(1-\frac{K}{2}\bar\zeta^2\right)\partial_{\bar\zeta}
\right),\\
\label{ZK}
\pmb{X}_3&=&\left(1+\frac{K}{2}\zeta^2\right)
\partial_\zeta+
\left(1+\frac{K}{2}\bar\zeta^2\right)\partial_{\bar\zeta},
\end{eqnarray}
closing in $\mathfrak{so}(3)$,  $\mathfrak{e}_2$ and $\mathfrak{so}(2,1)$ algebras for $K=+1, 0$ and $-1$ respectively.
The Carrollian structure is however richer as it hinges on the set $\left\{a_{ij}, b_i,\Omega\right\}$.
Hence, not all Riemannian isometries generated by a Killing field $\pmb{X}$ of $\mathscr{S}$ are necessarily promoted to Carrollian symmetries. For the latter, it is natural to further require the Carrollian vorticity be invariant:
\begin{equation}
\mathscr{L}_{\pmb{X}}\ast\varpi= \pmb{X}\left( \ast\varpi\right)=0. \label{Lie-var}
\end{equation}
Condition \eqref{Lie-var} is fulfilled for all fields $\pmb{X}_{A}$ ($A=1,2,3$) in  \eqref{XK}, \eqref{YK} and \eqref{ZK}, only as long as $a=\ell=0$, since $\ast\varpi=n$. Otherwise $\ast\varpi$ is non-constant and only 
$\pmb{X}_1=\text{i}\left(\zeta \partial_\zeta-\bar\zeta \partial_{\bar\zeta}\right)=\partial_\varPhi$ leaves it invariant. This is in line with the bulk isometry properties discussed earlier, while it provides a Carrollian-boundary manifestation of the rigidity theorem.

\subsection{Vorticity-free Carrollian fluid and the Ricci-flat Robinson--Trautman}

The zero-$k$ limit of the relativistic Robinson--Trautman fluid presented in  Sec. \eqref{sec:carfluid} 
 (Eqs. \eqref{holoantimet}--\eqref{RcurRT}) is in agreement with the direct Carrollian approach of Sec. \ref{sec:resum-riccif}.  Indeed, it is straightforward to check that the general formulas \eqref{c-Carrol}--\eqref{Psi-2-Carrol} give $c=0$ together with
\begin{equation}
\pmb{\chi}=\frac{\text{i}}{2}\left(\partial_{\zeta}K\text{d}\zeta-\partial_{\bar\zeta}K\text{d}\bar\zeta
\right),\quad \pmb{X}=\frac{\text{i}}{P^2}
\left(
\partial_{\zeta} \left(P^2\partial_t\partial_{\zeta}
\ln P
\right)\text{d}\zeta^2-
\partial_{\bar\zeta} \left(P^2\partial_t\partial_{\bar\zeta}
\ln P
\right)
\text{d}\bar\zeta^2 \right),
\end{equation}
 while $\psi_i=0=\Psi_{ij}$. These expressions satisfy \eqref{carEcot}--\eqref{carHcot}, 
and the duality relations \eqref{heat-resum-car-Q}, \eqref{visc-resum-car-sig} and 
\eqref{visc-resum-car-xi} lead to the friction  components of the energy--momentum tensor $Q_i$, $\Sigma_{ij}$ and $\Xi_{ij}$, precisely as they appear in  \eqref{QexpC-RT}, \eqref{sigexpC-RT}. 
The general hydrodynamic equations \eqref{heat-eq-carE}, \eqref{mom-eq-carG},  are solved with\footnote{Since $\pi_i$ is not related to the geometry by duality as the other friction and heat tensors,
it can \emph{a priori} assume any value. It is part of the Carrollian Robinson--Trautman fluid definition to set it to zero.}  
$\pi_i=0$ and $\varepsilon=\varepsilon(t)$ satisfying \eqref{carE2d}, \emph{i.e.}  Robinson--Trautman's  \eqref{RTcar}. 

Our goal is to present here the resummation of the derivative expansion \eqref{papaefresricf} into a Ricci-flat spacetime dual to the fluid at hand. The basic feature of the latter is that $b_i=0$ and $\Omega=1$, hence it is vorticity-free -- on top of being shearless. With these data, using \eqref{papaefresricf}, 
we find 
\begin{equation}
\label{papaefgentetr}
\text{d}s^2_{\text{RT}} =-2\text{d}t\left(\text{d}r +H \text{d}t\right)+2\frac{r^2}{P^2}\, \text{d}\zeta \text{d}\bar\zeta,
\end{equation}
where
\begin{equation}
\label{H}
2H = - 2r \partial_t \ln P+  K -\frac{2M(t)}{r},
\end{equation}
with $K= 2P^2 \partial_{\bar\zeta} \partial_\zeta\ln P$ the Gaussian curvature of \eqref{holoantimet}. 
This metric is Ricci-flat provided the energy density
 $\varepsilon(t) =\nicefrac{M(t)}{4\pi G}$ and the function $P=P(t,\zeta,\bar \zeta)$ 
satisfy \eqref{RTcar}. These are algebraically special spacetimes of all types, as opposed to the Kerr--Taub--NUT family studied earlier (Schwarzschild solution is common to these two families). Furthermore they never have twist ($\pmb{\psi}=\pmb{\Psi}=0$) and are generically asymptotically locally but not globally flat due to $\pmb{\chi}$ and $\pmb{X}$.

The specific Petrov type of Robinson--Trautman solutions is determined by analyzing the tensors \eqref{principal}, or \eqref{principal-holoQ} and  \eqref{principal-holosig} in holomorphic coordinates:
 \begin{equation}
 \varepsilon^+=\frac{M(t)}{4\pi G}
  ,\quad
 \label{principal-holoRT}
 \pmb{Q}^+
 =- \frac{\text{1}}{8\pi G}\partial_{\zeta}K \text{d}\zeta,
 \quad
 \pmb{\Sigma}^+
=- \frac{\text{1}}{4\pi GP^2}
\partial_{\zeta} \left(P^2\partial_t\partial_{\zeta}
\ln P
\right)\text{d}\zeta^2.
\end{equation}
We find the following classification (see \cite{Gath:2015nxa}):
\begin{description}
\item[II] generic;
\item[III] with $\varepsilon^+=0$ and $\nabla_i Q^{+i}=0$;
\item[N] with $\varepsilon^+=0$ and $ Q^+_i=0$;
\item[D] with $2 Q^+_{i}Q^{+}_{j}=3\varepsilon^+  \Sigma_{ij}^+$ and vanishing traceless part of $\nabla^{\hphantom{+}}_{(i} Q^+_{j)}$.
\end{description} 

\section{Conclusions}

The main message of our work is that starting with the standard AdS holography, there is a well-defined zero-cosmological-constant limit that relates asymptotically flat  spacetimes to Carrollian fluids living on their null boundaries. 

In order to unravel this relationship and make it operative for studying holographic duals, we used the derivative expansion. Originally designed for asymptotically anti-de Sitter spacetimes with cosmological constant $\Lambda=-3k^2$, this expansion provides their line element in terms of the conformal boundary data: a pseudo-Riemannian metric and a relativistic fluid. It is expressed in Eddington--Finkelstein coordinates, where the zero-$k$ limit is unambiguous: it maps the pseudo-Riemannian boundary $\mathscr{I}$ onto a Carrollian geometry $\mathbb{R}\times \mathscr{S}$, and the conformal relativistic fluid becomes Carrollian. 

The emergence of the conformal Carrollian symmetry in the Ricci-flat asymptotic is not a surprise, as we have extensively discussed in the introduction. In particular, the BMS group has been used for investigating the asymptotically flat dual dynamics. 
What is remarkable is the efficiency of the derivative expansion to implement the limiting procedure and deliver a genuine holographic relationship between Ricci-flat spacetimes and conformal Carrollian fluids. These are defined on $\mathscr{S}$ but their dynamics is rooted in $\mathbb{R}\times \mathscr{S}$. 

Even though proving that the derivative expansion is unconditionally well-behaved in the limit under consideration is still part of our agenda, we have demonstrated this property in the instance where it is resummable.  

The resummability of the derivative expansion has been studied in our earlier works about anti-de Sitter fluid/gravity correspondence. It has two features:
\begin{itemize}
\item the shear of the fluid congruence vanishes;
\item the heat current and the viscous stress tensor are determined from the Cotton current and stress tensor components via a transverse (with respect to the velocity) duality. 
\end{itemize}
The first considerably simplifies the expansion. Together with the second, it ultimately dictates  the structure of the bulk Weyl tensor, making the Einstein spacetime of special Petrov type.  The conservation of the energy--momentum tensor is the only requirement left for the bulk be Einstein. It  involves the energy density (\emph{i.e.} the only fluid observable left undetermined) and various geometric data in the form of partial differential equations (as is the Robinson--Trautman for the vorticity-free situation). 

This pattern survives the zero-$k$ limit,  taken in a frame where the relativistic fluid is at rest. The corresponding Carrollian fluid -- at rest \emph{by law} -- is required to be shearless, but has otherwise acceleration, vorticity and expansion. Since the fluid is at rest, these are geometric data, as are the descendants of the Cotton tensor used again to formulate the duality that determines the dissipative components of the Carrollian fluid.

The study of the Cotton tensor and its Carrollian limit is central in our analysis. In Carrollian geometry (conformal in the case under consideration) it 
opens the pandora box of the classification of curvature tensors, which we have marginally discussed here. Our observation is that the Cotton tensor grants the zero-$k$ limiting Carrollian geometry on $\mathscr{S}$ with a scalar, two vectors and two symmetric, traceless tensors, satisfying a set of identities inherited from the original conservation equation. 

In a similar fashion, the relativistic energy--momentum tensor descends in a scalar (the energy density), two heat currents and two viscous stress tensors. This doubling is suggested by that of the Cotton. The physics behind it is yet to be discovered, as it requires a microscopic approach to Carrollian fluids, missing at present. Irrespective of its microscopic origin, however, this is an essential result of our work, in contrast with previous attempts. Not only we can state that the fluid holographically dual to a Ricci-flat spacetime is neither relativistic, nor Galilean, but we can also exhibit for the actually Carrollian fluid the fundamental observables and the equations they obey.\footnote{ From this perspective, trying to design four-dimensional flat holography  using two-dimensional conformal   field theory described in terms of a conserved two-dimensional energy--momentum tensor \cite{Str0, Str1, Str2} looks inappropriate.} These are quite convoluted, and  whenever satisfied, the resummed metric is Ricci-flat.

Our analysis, amply illustrated by two distinct examples departing from Carrollian hydrodynamics and ending on widely used Ricci-flat spacetimes, raises many questions, which deserve a comprehensive survey. 

As already acknowledged, the Cotton Carrollian descendants enter the holographic reconstruction of a Ricci-flat spacetime, along with the energy--momentum data. It would be rewarding to 
explore the information stored in these objects, which may carry the boundary interpretation of the Bondi news tensor as well as of the asymptotic charges one can extract from the latter.

We should stress at this point that Cotton and energy--momentum data (and the charges they transport) play dual r\^oles. The nut and the mass provide the best paradigm of this statement. Altogether they raise the question on the thermodynamic interpretation of magnetic charges. Although we cannot propose a definite answer to this question, the tools of fluid/gravity holography (either AdS or flat) may turn helpful. This is tangible in the case of algebraically special Einstein solutions, where the underlying integrability conditions set a deep relationship between geometry and energy--momentum \emph{i.e.} between geometry and local thermodynamics.  To make this statement more concrete, observe the heat current as constructed using the integrability conditions, Eq. \eqref{heat-resum-car-Q}:
\begin{equation}
Q_i
=-
\frac{1}{16\pi G}\left(\hat{\mathscr{D}}_i\hat{\mathscr{K}}- \eta^j_{\hphantom{j}i} \hat{\mathscr{D}}_j\hat{\mathscr{A}}+4\ast \varpi  \eta^j_{\hphantom{j}i} \hat{\mathscr{R}}_j\right).
\nonumber
\end{equation}
In the absence of magnetic charges, only the first term is present and it is tempting to set
a relationship between the temperature and the gravito-electric curvature scalar $\hat{\mathscr{K}}$. This was precisely discussed in the AdS framework when studying the Robinson--Trautman relativistic fluid, in Ref. \cite{Ciambelli:2017wou}. Magnetic charges switch on the other terms, exhibiting natural thermodynamic potentials, again related with curvature components ($\hat{\mathscr{A}}$ and $ \hat{\mathscr{R}}_j$).

{ We would like to conclude with a remark. On the one hand, we have shown that the boundary fluids holographically dual to Ricci-flat spacetimes
are of Carrollian nature. On the other hand, the stretched horizon in the membrane paradigm seems to be rather described in terms of  Galilean hydrodynamics  \cite{Damourpaper, damour1979quelques, PrTh}. Whether and how these two pictures could been related is certainly worth refining.
}

\section*{Acknowledgements}

We would like to thank G. Barnich, G. Bossard, A. Campoleoni, S. Mahapatra, O. Miskovic, A. Mukhopadhyay, R. Olea and P. Tripathy  for valuable scientific exchanges. Marios Petropoulos would like to thank N. Banerjee for the \textsl{Indian Strings Meeting}, Pune, India, December 2016, P. Sundell, O. Miskovic and R. Olea for the \textsl{Primer Workshop de Geometr\' ia y F\' isica}, San Pedro de Atacama, Chile, May 2017,
and A. Sagnotti for the \textsl{Workshop on Future of Fundamental Physics} (within the  6th International Conference on New Frontiers in Physics -- ICNFP), Kolybari, Greece, August 2017, where many stimulating discussions on the topic of this work helped making progress. 
We thank each others home institutions for hospitality and financial support. 
This work was supported by the ANR-16-CE31-0004 contract \textsl{Black-dS-String} .

\appendix 

\section{Carrollian boundary geometry in holomorphic coordinates}\label{holo}

Using Carrollian diffeomorphisms \eqref{cardifs}, the metric \eqref{dmet} of the Carrollian geometry on the two-dimensional surface
$\mathscr{S}$ can be recast in conformally flat form,
\begin{equation}
\label{CF}
\text{d}\ell^2=\frac{2}{P^2}\text{d}\zeta\text{d}\bar\zeta
\end{equation}
with $P=P(t,\zeta,\bar \zeta)$ a real function, under the necessary and sufficient condition that the Carrollian shear $\xi_{ij}$ displayed in \eqref{carshexp-tempcon} vanishes. We will here assume that this holds and present a number of useful formulas for Carrollian and conformal Carrollian geometry.  These geometries carry two further pieces of data: $\Omega(t,\zeta,\bar \zeta)$ and  
\begin{equation}
\label{frame}
\pmb{b}=b_\zeta(t,\zeta, \bar \zeta)\, \text{d}\zeta+b_{\bar\zeta}(t,\zeta, \bar \zeta)\, \text{d}\bar\zeta
\end{equation}
with $b_{\bar\zeta}(t,\zeta, \bar \zeta)=\bar b_\zeta(t,\zeta, \bar \zeta)$. 
Our choice of orientation is inherited from the one adopted for the relativistic boundary (see footnote \ref{orient}) with  $a_{\zeta\bar\zeta}=\nicefrac{1}{P^2}$ is\footnote{This amounts to setting $\sqrt{a}=\nicefrac{\text{i}}{P^2}$ in coordinate frame and $\epsilon_{\zeta\bar\zeta}=-1$.}
\begin{equation}
\label{orientCF}
\eta_{\zeta\bar\zeta}=-\frac{\text{i}}{P^2}.
\end{equation}

The first-derivative Carrollian tensors are the acceleration \eqref{caracc}, the expansion \eqref{carshexp-tempcon} and 
the scalar vorticity \eqref{starvarpi}:
\begin{eqnarray}
&\varphi_{\zeta}
=\partial_t \dfrac{b_{\zeta}}{\Omega}+\hat\partial_{\zeta} \ln \Omega,\quad \varphi_{\bar\zeta}
=\partial_t \dfrac{b_{\bar\zeta}}{\Omega}+\hat\partial_{\bar\zeta} \ln \Omega,&
\label{acchol}
\\
&\theta =-\dfrac{2}{\Omega} \partial_t \ln P,
\quad 
\ast\varpi=\dfrac{\text{i}\Omega P^2}{2}\left(
\hat\partial_{\zeta}\dfrac{b_{\bar\zeta}}{\Omega}-\hat\partial_{\bar\zeta} \dfrac{ b_{\zeta}}{\Omega}
\right)&
\label{thetstarvarpihol}
\end{eqnarray}
with
\begin{equation}
\hat\partial_{\zeta} = \partial_\zeta+\frac{b_{\zeta}}{\Omega}\partial_t,\quad
\hat\partial_{\bar\zeta} = \partial_{\bar\zeta}+\frac{b_{\bar\zeta}}{\Omega}\partial_t.
\end{equation}  
Curvature scalars and vector are second-derivative (see \eqref{scalar},  \eqref{carriemanntime}):\footnote{We also quote for completeness (useful \emph{e.g.} in Eq. \eqref{CWscalarhol}):
$$
\hat K =K+ P^2 \left[ 
\partial_{\zeta} \frac{b_{\bar\zeta}}{\Omega}+
\partial_{\bar\zeta} \frac{b_{\zeta}}{\Omega}+
\partial_{t} \frac{b_{\zeta}b_{\bar\zeta}}{\Omega^2}
+2  \frac{b_{\bar\zeta}}{\Omega}\partial_{\zeta}
+2 \frac{b_{\zeta}}{\Omega} \partial_{\bar\zeta}
+2 \frac{b_{\zeta}b_{\bar\zeta}}{\Omega^2}\partial_{t}
 \right]\partial_t\ln P
$$
with $K= 2P^2 \partial_{\bar\zeta} \partial_\zeta\ln P$ 
the ordinary Gaussian curvature of the two-dimensional metric 
 \eqref{CF}.}
\begin{eqnarray}
&\hat K=P^2\left(\hat\partial_{\bar\zeta} 
 \hat\partial_{\zeta}
 + \hat\partial_{\zeta}\hat\partial_{\bar\zeta}
 \right)
\ln P,
\quad
\hat A=
\text{i}P^2\left(\hat\partial_{\bar\zeta} 
 \hat\partial_{\zeta}
 - \hat\partial_{\zeta}\hat\partial_{\bar\zeta}
 \right)
\ln P,&
\label{hatKA}
\\
&
\hat r_\zeta=\dfrac{1}{2}\hat\partial_{\zeta}\left(\dfrac{1}{\Omega}
\partial_t\ln P
\right),\quad
\hat r_{\bar\zeta}=\dfrac{1}{2}\hat\partial_{\bar\zeta}\left(\dfrac{1}{\Omega}
\partial_t\ln P
\right),
&
\label{hatr}
\end{eqnarray}
and we also quote:
\begin{eqnarray}
&
\ast \varphi=\text{i}P^2\left(
\hat\partial_{\zeta}\varphi_{\bar\zeta}-\hat\partial_{\bar\zeta} \varphi_{\zeta}
\right),
&
\\
&\hat\nabla_k\varphi^k= P^2\left[\hat\partial_\zeta \partial_t \frac{b_{\bar \zeta}}{\Omega}
+
\hat\partial_{\bar \zeta} \partial_t \frac{b_\zeta}{\Omega}
+\left(\hat\partial_{\zeta}\hat\partial_{\bar \zeta}+\hat\partial_{\bar \zeta}\hat\partial_\zeta
\right)\ln \Omega
\right].&
\end{eqnarray}

Regarding conformal Carrollian tensors we remind the weight-$2$ curvature scalars \eqref{CWscalar}:
\begin{equation}
\label{CWscalarhol}
\hat{\mathscr{K}}=\hat{K}+ \hat{\nabla}_k \varphi^k
,\quad \hat{\mathscr{A}}=  \hat{A}- \ast \varphi,
\end{equation}
and the  weight-$1$ curvature one-form \eqref{CWRvec}:
\begin{equation}
\hat{\mathscr{R}}_{\zeta}=\frac{1}{\Omega} \partial_{t}\varphi_{\zeta}-\frac{1}{2}\left(\hat \partial_{\zeta}+\varphi_{\zeta}\right)\theta, \quad \hat{\mathscr{R}}_{\bar\zeta}=\frac{1}{\Omega} \partial_{t}\varphi_{\bar\zeta}-\frac{1}{2}\left(\hat \partial_{\bar\zeta}+\varphi_{\bar\zeta}\right)\theta.
\label{CWRvechol}
\end{equation}
The three-derivative Cotton descendants displayed in \eqref{c-Carrol}--\eqref{Psi-2-Carrol} are a scalar
\begin{equation}
\label{c-Carrolhol}
c=\left(\hat{\mathscr{D}}_l\hat{\mathscr{D}}^l+2\hat{\mathscr{K}}
\right)\ast \varpi
\end{equation}  
of weight $3$ ($\ast \varpi$ is of weght $1$), two vectors
\begin{eqnarray}
\label{chi-f-Carrolhol}
&\chi_{\zeta}=\frac{\text{i}}{2}\hat{\mathscr{D}}_{\zeta}\hat{\mathscr{K}}+ \frac{1}{2} \hat{\mathscr{D}}_{\zeta}\hat{\mathscr{A}}-2\ast \varpi\hat{\mathscr{R}}_{\zeta},
\quad 
\chi_{\bar\zeta}=-\frac{\text{i}}{2}\hat{\mathscr{D}}_{\bar\zeta}\hat{\mathscr{K}}+ \frac{1}{2} \hat{\mathscr{D}}_{\bar\zeta}\hat{\mathscr{A}}-2\ast \varpi\hat{\mathscr{R}}_{\bar\zeta}
,
&
\\
\label{psi-f-Carrolhol}
&\psi_{\zeta}=3\text{i}\hat{\mathscr{D}}_{\zeta}\ast \varpi^2,
\quad \psi_{\bar\zeta}=-3\text{i}\hat{\mathscr{D}}_{\bar\zeta}\ast \varpi^2,&
\end{eqnarray}  
of weight $2$, and two symmetric and traceless tensors
\begin{eqnarray}
\label{X-2-Carrolhol}
&X_{\zeta\zeta}=\text{i}\hat{\mathscr{D}}_{\zeta}
\hat{\mathscr{R}}_{\zeta},\quad
X_{{\bar\zeta}{\bar\zeta}}=-\text{i}\hat{\mathscr{D}}_{{\bar\zeta}}
\hat{\mathscr{R}}_{{\bar\zeta}}
,&
\\
\label{Psi-2-Carrolhol}
&\Psi_{\zeta\zeta}=\hat{\mathscr{D}}_{\zeta} \hat{\mathscr{D}}_{\zeta}\ast \varpi ,
\quad 
\Psi_{{\bar\zeta}{\bar\zeta}}=\hat{\mathscr{D}}_{{\bar\zeta}} \hat{\mathscr{D}}_{{\bar\zeta}}\ast \varpi ,
&
\end{eqnarray}  
of weight $1$. Notice that in holomorphic coordinates a symmetric and traceless tensor $S_{ij}$ has only diagonal entries: $S_{\zeta\bar{\zeta}}=0=S_{\bar{\zeta}\zeta}$. 

We  also remind for convenience some expressions for the determination of Weyl--Carroll covariant derivatives. If $\Phi$ is a weight-$w$ scalar function
\begin{equation}
\label{CWs-Phi-hol}
\hat{\mathscr{D}}_{\zeta} \Phi=\hat\partial_{\zeta} \Phi +w \varphi_{\zeta} \Phi,\quad \hat{\mathscr{D}}_{\bar\zeta} \Phi=\hat\partial_{\bar\zeta} \Phi +w \varphi_{\bar\zeta} \Phi.
\end{equation}
For weight-$w$ form components $V_{\zeta} $ and $V_{\bar\zeta} $ the Weyl--Carroll derivatives read: 
\begin{eqnarray}
&\hat{\mathscr{D}}_{\zeta} V_{\zeta}=\hat\nabla_{\zeta} V_{\zeta} +(w+2) \varphi_{\zeta}V_{\zeta},\quad
\hat{\mathscr{D}}_{\bar\zeta} V_{\bar\zeta}=\hat\nabla_{\bar\zeta} V_{\bar\zeta} +(w+2) \varphi_{\bar\zeta}V_{\bar\zeta},&
\\
&\hat{\mathscr{D}}_{\zeta} V_{\bar\zeta}=\hat\nabla_{\zeta} V_{\bar\zeta} +w \varphi_{\zeta}V_{\bar\zeta},\quad
\hat{\mathscr{D}}_{\bar\zeta} V_{\zeta}=\hat\nabla_{\bar\zeta} V_{\zeta} +w \varphi_{\bar\zeta}V_{\zeta},&
\end{eqnarray}
while the Carrollian covariant derivatives are simply:
\begin{eqnarray}
&\hat\nabla_{\zeta} V_{\zeta} =\dfrac{1}{P^2}\hat\partial_{\zeta}\left(P^2
V_{\zeta}
\right),\quad
\hat\nabla_{\bar\zeta} V_{\bar\zeta} =\dfrac{1}{P^2}\hat\partial_{\bar\zeta}\left(P^2
V_{\bar\zeta}
\right),&
\\
&\hat\nabla_{\zeta} V_{\bar\zeta}= \hat\partial_{\zeta}V_{\bar\zeta},\quad
\hat\nabla_{\bar\zeta} V_{\zeta} =\hat\partial_{\bar\zeta}V_{\zeta}.&
\end{eqnarray}
Finally,
\begin{equation}
\label{weyllapl-scalhol}
\hat{\mathscr{D}}_k\hat{\mathscr{D}}^k \Phi=P^2\left(
 \hat\partial_{\zeta}\hat\partial_{\bar\zeta} \Phi +\hat\partial_{\bar\zeta} \hat\partial_{\zeta} \Phi 
+w \Phi\left( \hat\partial_{\zeta} \varphi_{\bar\zeta} 
+
 \hat\partial_{\bar\zeta} \varphi_\zeta
\right)
+2w\left(\varphi_{\zeta} \hat\partial_{\bar\zeta} \Phi +\varphi_{\bar\zeta} \hat\partial_{\zeta} \Phi 
+w\varphi_{\zeta} \varphi_{\bar\zeta} 
\Phi 
\right)\right).
\end{equation}

\end{document}